\documentclass[
reprint,
amsmath,amssymb,
aps,
prl,
showpacs,
twocolumn,
superscriptaddress,
floatfix,
preprintnumbers
]{revtex4-2}

\usepackage{bbm}
\usepackage{mathrsfs}
\usepackage{epsfig}

\usepackage{soul,xcolor}
\usepackage{graphicx}

\usepackage{amsfonts}
\usepackage{amsthm}
\usepackage[figuresright]{rotating}
\usepackage{amssymb}
\usepackage{amsmath}
\usepackage[shortlabels]{enumitem}
\usepackage{mathtools}
\usepackage{dcolumn}
\usepackage{physics}
\usepackage{float}
\usepackage{bm}
\usepackage{verbatim}
\usepackage{braket}
\usepackage[normalem]{ulem}
\usepackage[ruled,vlined,linesnumbered]{algorithm2e}
\usepackage{thm-restate}
\usepackage{setspace}
\usepackage[colorlinks,linkcolor=blue,anchorcolor=blue,citecolor=blue,urlcolor=blue]{hyperref}
\usepackage{stackengine}
\usepackage{svg}
\usepackage{qcircuit}

\usepackage{fancyhdr}

\DeclarePairedDelimiter{\parens}{(}{)}

\newcommand{\C}{\mathbb{C}}

\newcommand{\R}{\mathbb{R}}

\usepackage[capitalise]{cleveref}
\theoremstyle{plain}

\theoremstyle{definition}

\newcommand{\QSP}{\textrm{\normalfont QSP}}
\renewcommand{\O}[1]{O\left(#1\right)}
\usepackage{stmaryrd}

\usepackage{pdfpages} 
\usepackage{pgffor} 

\makeatletter
\AtBeginDocument{\let\LS@rot\@undefined}
\makeatother

\def\supplementfilename{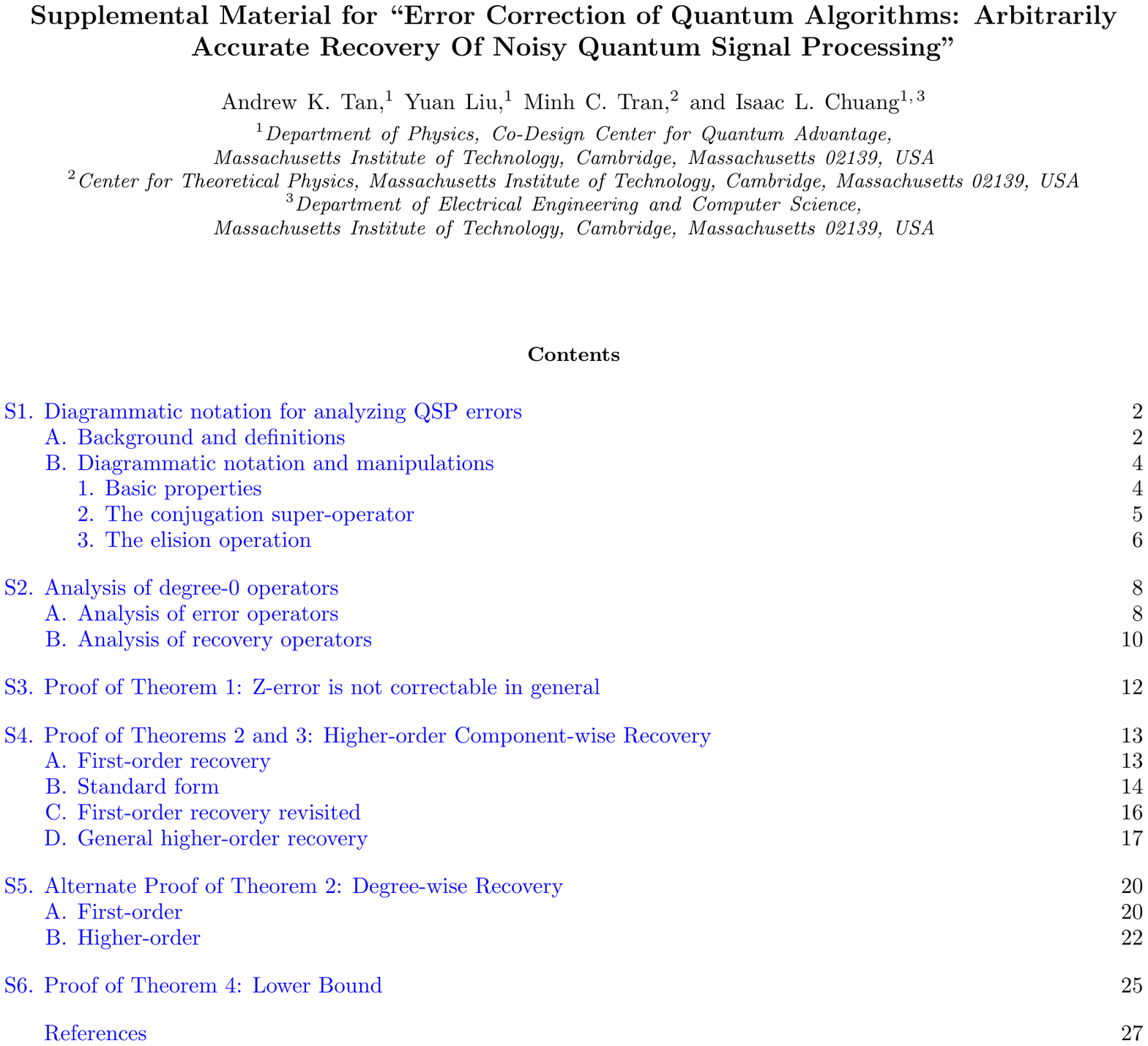}

\pdfximage{\supplementfilename}
\def\numbersupplementpages{\the\pdflastximagepages}

\begin{document}

\title{Error Correction of Quantum Algorithms: Arbitrarily Accurate Recovery Of Noisy Quantum Signal Processing}
\author{Andrew K. Tan}
  \email{aktan@mit.edu}
  \affiliation{Department of Physics, Co-Design Center for Quantum Advantage, Massachusetts Institute of Technology, Cambridge, Massachusetts 02139, USA}
\author{Yuan Liu}
  \email{yuanliu@mit.edu}
  \affiliation{Department of Physics, Co-Design Center for Quantum Advantage, Massachusetts Institute of Technology, Cambridge, Massachusetts 02139, USA}
\author{Minh C. Tran}
  \email{minhtran@mit.edu}
  \affiliation{Center for Theoretical Physics, Massachusetts Institute of Technology, Cambridge, Massachusetts 02139, USA}
\author{Isaac L. Chuang}
  \affiliation{Department of Physics, Co-Design Center for Quantum Advantage, Massachusetts Institute of Technology, Cambridge, Massachusetts 02139, USA}
  \affiliation{Department of Electrical Engineering and Computer Science, Massachusetts Institute of Technology, Cambridge, Massachusetts 02139, USA}

\vspace*{-.6in}

\preprint{MIT-CTP/5521}

\begin{abstract}
  The intrinsic probabilistic nature of quantum systems makes error correction or mitigation indispensable for quantum computation. 
  While current error-correcting strategies focus on correcting errors in quantum states or quantum gates, these fine-grained error-correction methods can incur significant overhead for quantum algorithms of increasing complexity.
  We present a first step in achieving error correction at the level of quantum algorithms by combining a unified perspective on modern quantum algorithms 
  via quantum signal processing (QSP). 
  An error model of under- or over-rotation of the signal processing operator parameterized by \(\epsilon < 1\) is introduced.
  It is shown that while Pauli \(Z\)-errors are not recoverable without additional resources, Pauli \(X\) and \(Y\) errors can be arbitrarily suppressed by coherently appending a noisy `recovery QSP.' 
  Furthermore, it is found that a recovery QSP of length \(O(2^k c^{k^2} d)\) is sufficient to correct any length-\(d\) QSP with \(c\) unique phases to \(k\textsuperscript{th}\)-order in error \(\epsilon\). 
  Allowing an additional assumption, a lower bound of \(\Omega(cd)\) is shown, which is tight for \(k = 1\), on the length of the recovery sequence.
  Our algorithmic-level error correction method is applied to Grover's fixed-point search algorithm as a demonstration.
\end{abstract}

\maketitle

\vspace*{-.1in}

Error correction and noise mitigation strategies are crucial for quantum computation.  Two important and distinct strategies have been suppression of systematic errors using composite pulses \cite{m:brownArbitrarilyAccurateComposite2004,m:lowOptimalArbitrarilyAccurate2014,m:levittCompositePulses1986} and correction of random errors using classical and quantum codes \cite{m:preskillReliableQuantumComputers1998,m:knillTheoryQuantumErrorcorrecting1997a,m:albertPerformanceStructureSinglemode2018}. Both of them are fine-grained approaches that aim to make more perfect gates from imperfect ones.  Remarkably, modern classical computers almost completely eschew such fine-grained error correction and, in their place, employ more efficient strategies which correct or control errors at the level of software and algorithms~\cite{koren2020fault}.

Envisioning similar algorithmic-level error correction for quantum computation is challenging;  simply repeat-until-success strategies (e.g. error mitigation through detection and post-selection) would incur exponential overhead~\cite{m:wangCanErrorMitigation2021a,m:quekExponentiallyTighterBounds2022c}.  
Given recent progress providing a unifying perspective on modern quantum algorithms through quantum signal processing (QSP) \cite{m:lowMethodologyResonantEquiangular2016,m:lowHamiltonianSimulationQubitization2019}, its generalization---quantum singular value transformation (QSVT) \cite{m:gilyenQuantumSingularValue2019, m:martynGrandUnificationQuantum2021}, and sophisticated new understandings of error propagation in Lie product formulas \cite{m:childsTheoryTrotterError2021b}, might there be a new strategy combining these insights to enable error correction at the level of quantum algorithms?
The defining feature of such algorithm-level error correction (ALEC) techniques is the design of subroutines that allow gate-level errors to cancel; therefore requiring a sophisticated understanding of how gate-level errors propagate to the output of algorithms. Such error propagation is typically accomplished via analysis of non-commutative algebra \cite{m:childsTheoryTrotterError2021b}.

Here, we demonstrate a first example of ALEC under the simple noise model of a consistent multiplicative under- or over-rotation in the QSP operators by a fixed but unknown noise parameter \(\epsilon\). 
Given such a noisy QSP sequence, we construct a ``recovery sequence''---a complementary QSP sequence subjected to the same noise and to be appended to the original sequence---that corrects the error up to the \(k\textsuperscript{th}\) order in \(\epsilon\), for an arbitrary \(k \ge 1\). Importantly, our method fits squarely within the standard QSP model and does not require any additional resources (e.g. additional signal processing axes). Our construction is akin to composite pulses in that the quantum ALEC works despite being blind to \(\epsilon\), but the fact that the recovery sequence must additionally work for all values of the \emph{unknown} input ``signal'' rotation marks a departure from the composite pulse setting and makes recovery substantially more challenging. In addition, our construction is fully coherent and deterministic, which differs from known error mitigation techniques \cite{m:temmeErrorMitigationShortDepth2017,m:hugginsVirtualDistillationQuantum2021a,m:heZeronoiseExtrapolationQuantumgate2020,m:bergProbabilisticErrorCancellation2022}.

\begin{figure}[t]
  \includegraphics[width=0.49\textwidth]{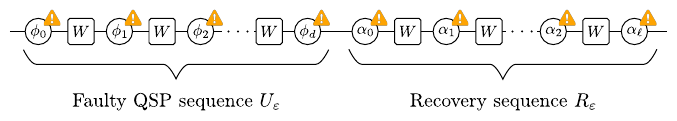}
  \caption{Many important quantum algorithms can be reformulated as transformation on some ``signal," which can be implemented by a QSP sequence consisting of queries to a signal operator \(W\) and rotations parameterized by phase angles \(\{ \phi \}\)'s to manipulate the signal. 
  Imperfections in these rotations result in errors in the quantum algorithms. 
  We recover the correct outputs of the algorithms by appending recovery sequences \(R_\epsilon\).
  The phase angles \(\{ \alpha\}\)'s of the recovery sequences depend only on the angles \(\{\phi\}\)'s of the bare sequences.
  }
\label{fig:demo}
\end{figure}

The rest of the paper is organized as follows. 
After reviewing the QSP framework for quantum algorithms, we introduce the error model and explore the feasibility of error correction using only noisy QSP operators.
We present our main results, including \cref{thm:no-z-recovery} on the impossibility of the most general form of recovery.
This no-go theorem  motivates a restricted form of recovery which we show is possible in \cref{thm:existence-of-recovery-sequence}.
Our constructive proof allows us to place upper-bounds on the resource requirements for recovery which we present in \cref{thm:recovery-length-upper}.
Next, allowing an additional assumption, we prove the optimality of our construction \cref{thm:recovery-length-lower}.
These results on QSP can be immediately lifted to QSVT. 
As an example, we demonstrate an application of our algorithmic error correction strategy to the Grover's search algorithm. 
Finally, we conclude the paper with a discussion on implications of our results. 

\emph{Framework \& Main results.---}
  We consider a computation where the input \(\theta\) is encoded in a \emph{signal} operator 
  \begin{align} 
    W(\theta) \equiv e^{i\theta X}  =  \begin{pmatrix}
      \cos \theta & i \sin \theta \\
      i \sin\theta & \cos \theta
    \end{pmatrix}
  \end{align}
  and the quantum algorithm is a length-$d$ QSP sequence of the form
  \begin{align} 
    U(\theta) = \QSP(\theta;\vec\phi) \equiv  e^{i \phi_0 Z} \prod_{j = 1}^d W(\theta) e^{i \phi_j Z}, 
  \end{align}
  where \(X, Z\) are the Pauli matrices and \(\vec \phi = (\phi_0,\dots,\phi_d) \in \R^{d+1}\) are the QSP phase angles.
  The rotations around the \(Z\) axis ``process'' the signal, i.e. they transform entries of \(U\) into functions of \(\cos \theta\).
  In particular, given a polynomial \(P(\cos\theta)\) of degree at most \(d\) in \(\cos\theta\) satisfying certain conditions, Remez-type algorithms \cite{m:fraserSurveyMethodsComputing1965} guarantee the existence of a set of phase angles \(\vec \phi\) such that the entry \(\braket{0 | U | 0}\) is exactly \(P(\cos\theta)\).
  Therefore, designing quantum algorithms, i.e. computing a desired function \(f(\cos\theta)\), reduces to finding a good polynomial approximation \(P(\cos\theta) \approx f(\cos\theta)\).

  Generally, increasing the length $d$ of the QSP sequence enables better approximation. 
  However, due to accumulating experimental errors, the longer the QSP sequence, the further the computation deviates from the ideal $P(\cos\theta)$.
  We make the first step towards correcting these errors on the algorithmic level by constructively proving the existence of ALEC for a simple error model.

  In this error model, we assume that the signal processing operators under- or over-rotate by a fixed multiplicative factor $\epsilon$: $\phi \mapsto \phi(1+\epsilon)$ for all $\phi$ (Fig. \ref{fig:demo}).
  While \(\epsilon\) is unknown a priori, we assume that it is constant throughout the application of the sequence and that it is small \(\epsilon \ll 1\) so that we may expand errors in orders of \(\epsilon\).
  Such an error may be due to imperfections on the hardware control or non-optimality of classical computation of the QSVT processing phases. 
  A natural question to ask is how one should define recovery and if it is possible, given access only to such noisy signal processing rotations.

\emph{Impossibility of general recovery.---}
  In the most general form, one may ask given a noisy QSP \(U_\epsilon(\theta) = \QSP(\theta; \vec{\phi})\), if there exists another noisy QSP sequence \(U_\epsilon'(\theta) = \QSP(\theta;\vec \phi')\), characterized by phase angles \(\vec \phi'\) that depend on \(\vec \phi\), such that
  \begin{align} 
      \bra{0}U_\epsilon'\ket{0} = \bra{0} U_0 \ket{0} + \O{\epsilon^{k+1}}
  \end{align}
  reproduces the target polynomial \(P(\cos\theta) = \bra{0} U_0 \ket{0}\) up to the \(k\textsuperscript{th}\)-order in \(\epsilon\), where \(U_0 \equiv U_{\epsilon = 0}\).
  If this construction were possible, it would allow us to build a less noisy \(Z\) rotation, i.e. one that simply suppresses the over- or under-rotations in the phase angles.
  Using these rotations in place of the original, our recovered sequence would only be a constant factor longer than the original.
   
  Unfortunately, our first result is the impossibility of such a construction:
  \begin{restatable}[No correction of Z-error]{theorem}{nozcorrection}
  \label{thm:no-z-recovery}
    Let \(U_\epsilon\) be a length-\(d\) noisy QSP unitary parameterized by \(
    \parens{\phi_0, \ldots, \phi_d} \in \R^{d+1}\).
    For general phases \(\phi_i\), no noisy QSP unitary \(U_\epsilon'\) exists such that for any \(k \ge 1\),
    \begin{equation}
    \label{eq:z-recovery-condition}
      \bra{0} U_\epsilon' \ket{0} = \bra{0} U_0 \ket{0} + O(\epsilon^{k + 1}).
    \end{equation}
  \end{restatable}
  The result comes precisely from the fact that we are unable to build less noisy \(Z\) rotations from more noisy \(Z\) rotations in the noisy QSP framework.
  Consequently, it is only possible to contruct \(U_\epsilon'\) such that \(\braket{0 | U_\epsilon' | 0}\) approximates \(\braket{0 | U_0 | 0}\) up to a phase that may depend on \(\theta\). 
  We leave the detailed proof of \cref{thm:no-z-recovery} to the Supplemental Material (SM)~\cite{sm}.

\emph{\(XY\)-recovery.---}
  In light of this impossibility result, we search for a weaker, but nonetheless useful, definition of recovery.
  The primary challenge with the general form of recovery is the recovery of the \(Z\) error, which correspond to a global phase in the processed signal.
  In many applications where the QSP ancilla qubit is measured in the \(Z\) basis, only the measurement outcome probability \(\abs*{\braket{0 | U_\epsilon' | 0}}^2 \approx \abs*{P(\cos\theta)}^2\) is important and, thus, the global phase due to the \(Z\) error becomes irrelevant. 

  Our second result is a theorem establishing the existence of such a sequence \(U_\epsilon'\):
  \begin{restatable}[Recoverability]{theorem}{existence}
  \label{thm:existence-of-recovery-sequence}
    Given any noisy QSP operator \(U_\epsilon(\theta)\) and an integer \(k \geq 1\), there exists a recovery sequence \(R_\epsilon(\theta)\) satisfying
    \begin{equation}
      \abs{\braket{0 | U_\epsilon R_\epsilon | 0}}^2 = \abs*{\braket{0 | U_0 | 0}}^2 + O(\epsilon^{k + 1})
      \label{eq:thm1}
    \end{equation}
    for all \(\theta\).
  \end{restatable}
  To construct the first-order recovery operator, consider a length-\(d\) noisy QSP operator \(U_\epsilon\) parameterized by \((\phi_0, \dots, \phi_d)\) which we can write as
  \begin{align}
    \begin{split}
      U_\epsilon = U_0 [I + i \epsilon (x(\theta) \sin\theta X &+ y(\theta) \sin\theta Y \\
      &+ z(\theta) \cos\theta Z) + O(\epsilon^2)],
    \end{split}
  \end{align}
  for polynomials \(x, y, z \in \C[\cos\theta]\).
  Note that \(x\) and \(y\), which we call the error polynomials, are odd with degree at most \((2s-1)\), \(s \le d\).
  We can find a length-\((2s)\) QSP \(V^{(s)}_\epsilon\) parameterized by \((-\alpha^{(s)}_d, \dots, -\alpha^{(s)}_0, \pi/2, \alpha^{(s)}_0, \dots, \alpha^{(s)}_d)\) with error polynomials \(x_s\) and \(y_s\) of degree-\((2s - 1)\) such that \(x(\theta) + \beta_s x_s(\theta)\) and \(y(\theta) + \beta_s y_s(\theta)\) are polynomials of degree-\((2s-3)\) for \(\beta_s \in \R\).
  We can choose \(\delta_s \in \R\) such that by counter-rotating, \(U_\epsilon e^{i \delta_s (1 + \epsilon) Z} V^{(s)}_\epsilon e^{-2 i \delta_s (1 + \epsilon) Z} V^{(s)}_\epsilon e^{i \delta_s (1 + \epsilon) Z}\) has error polynomials \(x + \beta_s x_s\) and \(y + \beta_s y_s\).
  The entire first-order recovery operator is constructed in this way, each time reducing the degree of the error polynomials by two.
  The construction of higher-order recovery operators proceeds analogously and is done order-by order.
  A detailed discussion of the higher-order recovery operators can be found in the SM~\cite{sm}.

  \emph{Performance.---}To quantify the feasibility of error correcting a QSP sequence, we analyze the length of the recovery sequence in our construction.
  From the construction above, we see that to correct the degree-\((2s)\) term, we need to append a length \(\Theta(s)\) recovery operator.
  Since \(s \le d\), we require a \(O(d^2)\) operator for first-order recovery.
  For QSPs with phase degeneracies (i.e. phases differing by an integer multiple of \(2\pi\)), and therefore fewer real degrees of freedom, we can further economize by combining counter-rotation steps.
  In this way, the first-order recovery requires only \(O(cd)\) phases for a length-\(d\) QSP with \(c\) unique phases.
  Analogous savings can be found at higher-order resulting in the following result:
  \begin{restatable}[Upper bound on recovery length]{theorem}{upperbound}
  \label{thm:recovery-length-upper}
    Given any noisy QSP operator \(U_\epsilon(\theta)\) of length \(d\) with \(c\) distinct phases (up to factors of \(2\pi\)) and an integer \(k \geq 1\), there exists a recovery sequence \(R_\epsilon(\theta)\) satisfying \cref{thm:existence-of-recovery-sequence} with length at most \(O(2^k c^{k^2} d)\).
  \end{restatable}

\begin{figure}[t]
    \includegraphics[width=0.40\textwidth]{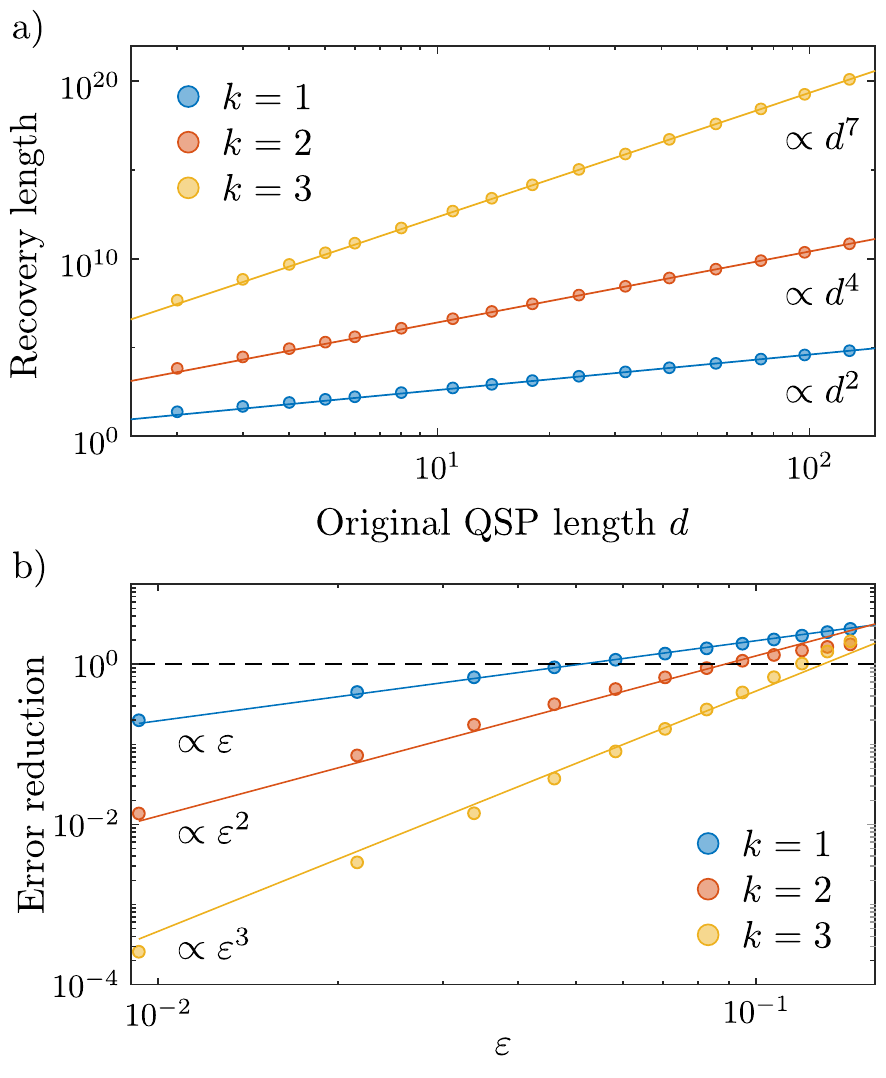}
    \caption{a) The exact length of our recovery sequence for original QSP sequences of different length \(d\) (scatter points) and \(c = d\) unique phases; solid lines show asymptotic scaling with \(d\). Upper bound with respect to \(d\) and \(k\) given by \cref{thm:recovery-length-upper}.
    b) The ratio between the error after and before applying the recovery sequence to a randomly generated QSP sequence as a function of \(\epsilon\) for different \(k\) (scatter points). The dashed horizontal line means no error reduction. We also plot functions proportional to \(\epsilon, \epsilon^2\), and \(\epsilon^3\) for reference.}
  \label{fig:benchmark}
  \end{figure}

  We plot the exact length of the recovery sequence as a function of \(d\) in \cref{fig:benchmark}a. 
  We note that, while the length increases exponentially with \(k\), the desired correction order \(k\) is usually a fixed constant.
  In particular, to correct the first-order error in \(\epsilon\) (\(k = 1\)), the length of the recovery sequence simply scales in the worse case (\(c \propto d\)) quadratically with the length of the bare sequence \(d\).

\emph{The benefit of degenerate phases.---}
  Interestingly, the length of the recovery operator in \cref{thm:recovery-length-upper} is dominated by \(c^{k^2}\) for large \(k\) which is dependent on the number of \emph{unique} phases \(c\).
  This suggests the number of unique phases may be a more useful measure of complexity than length in some cases.
  Indeed, in QSP constructions to quantum algorithms, it is not uncommon for the number of unique phases to scale sub-linearly with \(d\), as multiple signal processing phases may be identical \cite{m:yoderFixedPointQuantumSearch2014,m:groverQuantumComputersCan1998,m:martynGrandUnificationQuantum2021}; Grover's search algorithm is a notable example with one unique phase and is discussed further below.

\emph{Optimality. ---}
  Next, we prove a complementary theorem on a lower bound for the length of a recovery sequence to first-order in \(\epsilon\).
  \begin{restatable}[Lower bound on recovery length]{theorem}{lowerbound}
  \label{thm:recovery-length-lower}
    There exists a length-\(d\) QSP sequence \(U_\epsilon\) such that for any \(XY\) recovery sequence \(R_\epsilon\) of order \(k \ge 1\) satisfying 
    \begin{equation}
      U_0^\dagger U_\epsilon R_\epsilon = I + \epsilon f(a) e^{i \frac{\pi}{2} Z} + O(\epsilon^2),
    \end{equation}
    for function \(f(a) = O(a^0)\), \(R_\epsilon\) has length \(\Omega(d^2)\).
  \end{restatable}
  The assumption on the first-order \(Z\) component \(f(a) = O(a^0)\) in \cref{thm:recovery-length-lower} is required for a technical reason and effectively forces a unique choice of \(\eta^{(s)}_i\) in our first-order recovery operator, and can also be seen as a desire to limit the complexity of the recovery sequence.
  We conjecture that this assumption can be removed.
  Comparing \cref{thm:recovery-length-lower} to \cref{thm:recovery-length-upper} reveals that our construction is optimal for \(k = 1\).
  However, there is currently a gap between the lower bound and the construction for recovery beyond the first order \((k > 1)\), leaving room for future improvements.

  Closing these gaps between the upper and lower bounds has an important implication for quantum computation, given that even the \(O(d^2)\) scaling in our construction for first-order error correction would negate all quantum advantage (quadratic speedup) in most fixed-point quantum unstructured search \cite{yoder2014fixed}. 

\begin{figure}[b]
    \includegraphics[width=0.45\textwidth]{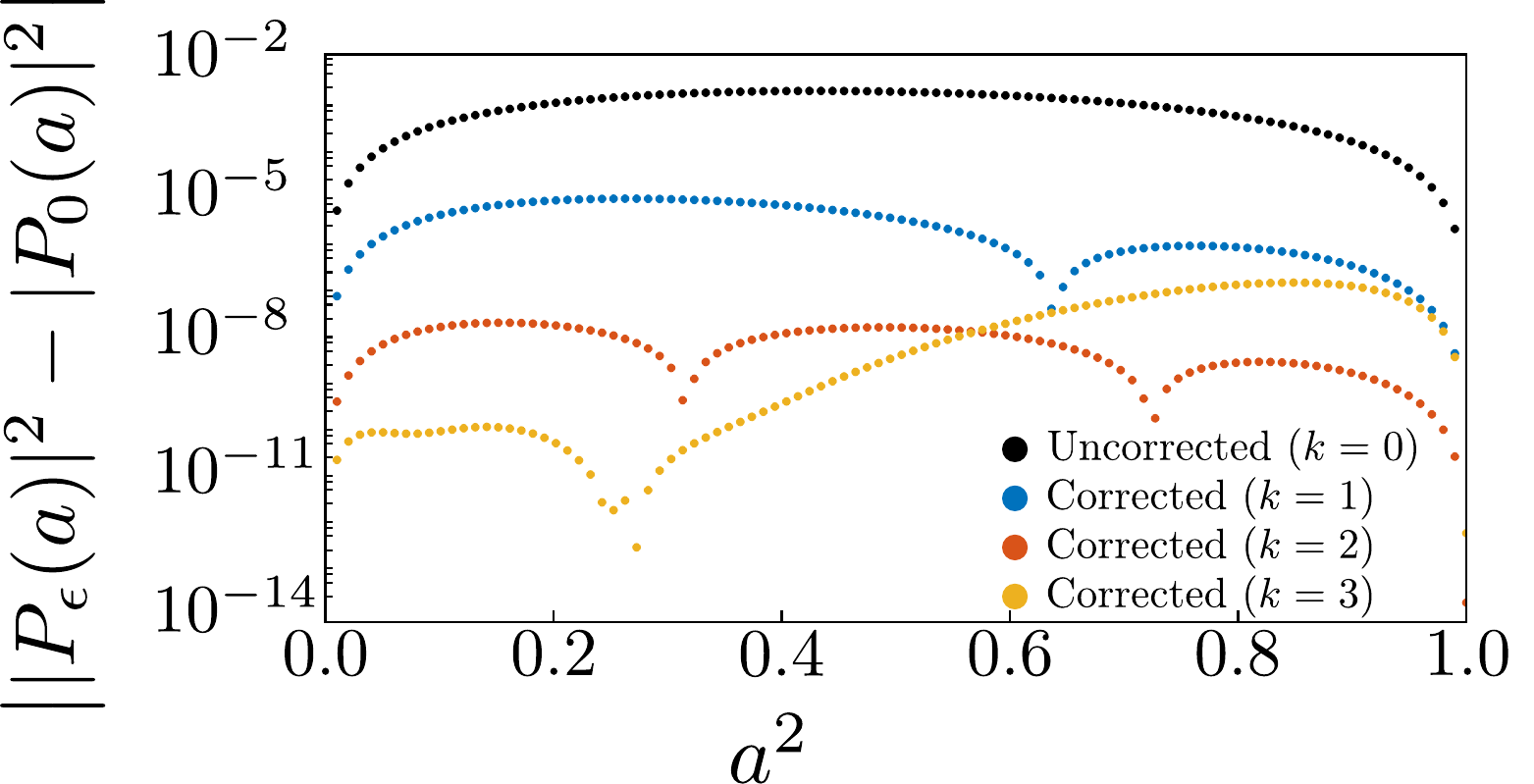}
    \caption{Deviation of success probability from ideal QSP function of a modified Grover's fixed-point search algorithm as a function of initial success probability demonstrated for the length-\(3\) QSP parameterized by \(\{\pi/3, \pi/3, \pi/3, \pi/3\}\). Various orders of error correction are shown at \(\epsilon = 0.001\). Note that for some values of \(a\), the \(k=3\) error exceeds the \(k=2\) error at this value of \(\epsilon\) due to the presence of large constant factors.}
  \label{fig:search}
  \end{figure}

  \emph{Application to Grover's Search as an example.---}We apply ALEC to a representative quantum algorithm, a modified version Grover's fixed-point search algorithm \cite{m:groverFixedPointQuantumSearch2005,yoder2014fixed} with a single unique phase, to demonstrate its utility. 
  By construction, the Grover fixed-point search algorithm amplifies a given amplitude \(a\) by substituting the reflection operator with \(\pm\pi/3\) rotations about the selected state, thus avoiding the souffl\'e problem of the non-fixed point Grover algorithm \cite{m:groverQuantumMechanicsHelps1997};
  while the original sequence is defined recursively, we study a modified length-\(3\) version with one unique phase.
  Fig. \ref{fig:search} shows the deviation of the noisy success probability \(|P_\epsilon(a)|^2\) from the noiseless \(|P_0(a)|^2\) as a function of the squared input overlap \(a = \cos^2\theta\) for varying levels of error correction at fixed \(\epsilon\). 
  It can be seen that the error is systematically reduced as ALEC is applied to correct higher and higher order errors, though for the plotted \(\epsilon = 0.001\), the error for the \(k=3\) sequence exceeds that of the \(k=2\) sequence for some values of \(a\) due to the presence of large constant factors.
  Nevertheless, for small enough \(\epsilon\), the error is suppressed, demonstrating the success of our ALEC theory.
  Explicit phases for the first-order recovery operator can be found in SM~\cite{sm}. 

\emph{Error threshold.---}
  Similar to standard quantum error correction, there is an error threshold beyond which the recovery sequence would instead add more error to the original QSP sequence.
  In \cref{fig:benchmark}b, we plot the error after applying the recovery sequences to a randomly generated QSP sequences as a function of \(\epsilon\).
  At \(k = 1\), for example, the recovery sequence becomes ineffective when \(\epsilon\) increases beyond \(\epsilon^*_1 = 0.05\).
  However, unlike standard quantum error correction, we can increase the threshold \(\epsilon^*_k\) and, thus, increase the error tolerance of the QSP sequence by increasing the order \(k\) in our construction. 
  An interesting open question is whether the error threshold \(\epsilon^*_k\) can be made arbitrarily large, i.e. \(\lim_{k\rightarrow \infty}\epsilon^*_k \rightarrow \infty\). 
  A positive answer to the question would provide a strong motivation for quantum devices based on the QSP architecture.

  \emph{Generalization to QSVT.---}
  The constructed recovery sequence in QSP can be immediately applied to the case of QSVT to simultaneously correct errors in all singular value subspaces due to qubitization \cite{m:lowHamiltonianSimulationQubitization2019}. More concretely, for a block-encoded operator \(A\) of dimension \(M \times M\), each singular value is associated with an SU(2) space which is transformed independently by the same polynomial transformation controlled by the QSVT phases. 
  Since our construction of the recovery sequence does not depend on the signal operator (the value of \(\theta\) in previous discussion, and in the QSVT case the signal being transformed is the singular value), the same recovery sequence works for all \(M\) singular values in each \(SU(2)\) subspace.  
  By virtue of the unification of major quantum algorithms provided by QSVT\cite{m:martynGrandUnificationQuantum2021}, this quantum ALEC construction can correct these kinds of errors arising in quantum search, simulation, and factoring algorithms.

\emph{Conclusions and Outlook.---}
  While it is already technically challenging to construct recovery sequences given the simple deterministic error model that we consider, it is absolutely crucial in the future to analyze the recovery sequence in the presence of an extensive source of random errors.
  These random errors typically introduce entropy into the quantum circuit and often arise in various quantum algorithms and physical devices. 

  Moreover, when our quantum algorithmic-level error-correction strategy is generalized to QSVT, two additional sources of errors should be considered: i) the projector-controlled-NOT operations, and  ii) the block-encoded signal unitaries. 
  While error i) may come from our inherent ignorance on the location of the subsystem of interest, resulting in erroneous projector. Error ii) can arise from approximations made in block-encoding the operator of interest. In general, these errors may be correlated to each other, and whether they are correctable depends on the specifics of the error models. Whether the recovery sequences exist in general and, if so, how their lengths scale with $d$ play an important role in harnessing quantum advantage from QSVT algorithms.

  In addition, to investigate more complicated error sources, we anticipate further development of the diagrammatic perturbative expansion\cite{sm} used in the present work as a formal tool to analyze error propagation in QSP. We hope such diagrammatic geometry can serve as a complimentary picture that eases future development of new algorithmic-level error correction strategies.

  Finally, we note that quantum ALEC is complementary to the standard state-level quantum error correcting (QEC) codes.
  While QEC codes protect quantum states by encoding information into an extended Hilbert \emph{space}, ALEC protects information by introducing redundancies in \emph{time}.
  Whereas standard QEC moves entropy into ancillary Hilbert spaces, one can view our construction as moving errors into the \(Z\) component.
  This also proves a limitation of our method as the $Z$-error can be important for situations when the QSP sequence need to be coherently concatenated with another quantum circuit \cite{m:martynEfficientFullycoherentQuantum2023}. 
  A combination of ALEC with standard QEC codes would provide a unified framework for fault-tolerant quantum computation with a tunable trade-off between space (number of qubits) and time (gate depth).

\bibliography{zotero-generated,ref}



\section*{Supplementary material}

## S1. Diagrammatic notation for analyzing QSP errors

In this Appendix, we provide a brief review of the QSP conventions and notations used in this paper. Additionally, we introduce a general diagrammatic notation which is useful for visualizing the perturbative analysis of our error model and more broadly for visualizing QSP operators.

### A. Background and definitions

**Definition S1.1.** A length-$d$ QSP operator is parameterized by phases $\vec{\phi} = (\phi_0, \ldots, \phi_d) \in \mathbb{R}^{d+1}$,

$$U(\theta; \vec{\phi}) = \text{QSP}(\theta; \vec{\phi}) \equiv e^{i\phi_0 Z} \prod_{j=1}^{d} W(\theta) e^{i\phi_j Z}, \tag{S1}$$

where

$$W(\theta) \equiv e^{i\theta X} = \begin{pmatrix} \cos\theta & i\sin\theta \\ i\sin\theta & \cos\theta \end{pmatrix} \tag{S2}$$

and $X$, $Y$, and $Z$ are the Pauli matrices. A general length-$d$ QSP sequence takes the form

$$U(\theta) = \begin{pmatrix} P(a) & iQ(a)\sqrt{1-a^2} \\ iQ^*(a)\sqrt{1-a^2} & P^*(a) \end{pmatrix}, \tag{S3}$$

where $a \equiv \cos\theta$ and $P, Q$ are polynomials of degrees at most $d$ [S1]. We write $U = [\![P_U, Q_U]\!]$ as a shorthand for Eq. (S3), dropping the subscripts when the QSP unitary is clear by context.

We consider an error model where the signal processing $Z$ rotation consistently under- or over-rotates by taking phases $\phi \mapsto \phi(1 + \epsilon)$. While $\epsilon$ is unknown a priori, we assume that is constant throughout the application of the QSP sequence and that it is small $\epsilon \ll 1$, allowing us to perform our error analysis perturbatively in orders of $\epsilon$. It is with this knowledge that we making the following definitions:

**Definition S1.2** (Canonical expansion). We say that an operator $U_\epsilon(\theta)$ admits a canonical expansion if we can write $U_\epsilon = w_\epsilon(\theta) I + i[x_\epsilon(\theta) X + y_\epsilon(\theta) Y + z_\epsilon(\theta) Z]$, for functions $w_\epsilon, x_\epsilon, y_\epsilon, z_\epsilon$ of the form

$$w_\epsilon(\theta) = \cos^2\theta \sum_{k=0}^{\infty} \epsilon^k \sum_{j=-1}^{\infty} \mathcal{P}_j^{(0,k)} \cos^{2j}(\theta), \tag{S4}$$

$$x_\epsilon(\theta) = \sin(2\theta) \sum_{k=0}^{\infty} \epsilon^k \sum_{j=0}^{\infty} \mathcal{P}_j^{(x,k)} \cos^{2j}(\theta), \tag{S5}$$

$$y_\epsilon(\theta) = \sin(2\theta) \sum_{k=0}^{\infty} \epsilon^k \sum_{j=0}^{\infty} \mathcal{P}_j^{(y,k)} \cos^{2j}(\theta), \tag{S6}$$

$$z_\epsilon(\theta) = \cos^2\theta \sum_{k=0}^{\infty} \epsilon^k \sum_{j=-1}^{\infty} \mathcal{P}_j^{(z,k)} \cos^{2j}(\theta), \tag{S7}$$

and $\mathcal{P}_j^{(\sigma,k)} \in \mathbb{R}$ for all $\sigma \in \{0, x, y, z\}$ and $j, k \in \mathbb{Z}$. We call $\mathcal{P}$ the *canonical profile* of $U_\epsilon$. For convenience, we allow $j \in \mathbb{Z}$ and define $\mathcal{P}_j^{(x,k)} = \mathcal{P}_j^{(y,k)} = \mathcal{P}_{j-1}^{(z,k)} = 0$ for all $j < 0$ and $k$.

Our parameterization, particularly the choice of factoring out $\sin(2\theta)$ from the $X$ and $Y$ components, $\cos^2\theta$ from the $Z$ component, and starting the sum of the $Z$ component at $j = -1$, is tailored to the error operators which we study in Section S2. Note that for unitary $U_\epsilon$, the functions satisfy the following completeness relationship $w_\epsilon(\theta)^2 + x_\epsilon(\theta)^2 + y_\epsilon(\theta)^2 + z_\epsilon(\theta)^2 = 1$ which holds for all $\theta$ and $\epsilon$.

The following Lemma demonstrates a class of operators, namely linear combinations of noiseless even-length QSP unitaries, that admit canonical expansions.

**Lemma S1** (QSP operators of even length exhibit canonical expansion). *Let $U_0(\theta) = \mathrm{QSP}(\theta; \vec{\phi})$ be a noiseless QSP unitary of length-2d, then $U_0(\theta)$ admits a canonical expansion at zeroth order in $\epsilon$. Additionally, $\mathcal{P}_j^{(\sigma,k)} = 0$ for all $k > 0$ and for $k = 0$ with $\sigma \in \{0, x, y, z\}$, and $j \geq d$.*

*Proof.* A QSP unitary of even length $2d$ can be written in the form Eq. (S3) with polynomials $P, Q \in \mathbb{C}[a]$ where $P$ has degree at most $2d$ and is of even parity in $\cos\theta$; and $Q$ has degree at most $2d - 1$ and is of odd parity [S1]. Writing the polynomials as

$$P(\cos\theta) = \sum_{j=0}^{d} p_{2j} \cos^{2j}(\theta), \tag{S8}$$

$$Q(\cos\theta) = \sum_{j=0}^{d-1} q_{2j+1} \cos^{2j+1}(\theta), \tag{S9}$$

we have

$$w_0(\theta) = \mathrm{Re}[P(\cos\theta)] = \cos^2\theta \sum_{j=-1}^{d-1} \mathrm{Re}[p_{2j}]\cos^{2j}(\theta), \tag{S10}$$

$$x_0(\theta) = \sin\theta\,\mathrm{Re}[Q(\cos\theta)] = \sin(2\theta)\sum_{j=0}^{d-1} \frac{1}{2}\mathrm{Re}[q_{2j+1}]\cos^{2j}(\theta), \tag{S11}$$

$$y_0(\theta) = -\sin\theta\,\mathrm{Im}[Q(\cos\theta)] = -\sin(2\theta)\sum_{j=0}^{d-1} \frac{1}{2}\mathrm{Im}[q_{2j+1}]\cos^{2j}(\theta), \tag{S12}$$

$$z_0(\theta) = \mathrm{Im}[P(\cos\theta)] = \cos^2\theta \sum_{j=-1}^{d-1} \mathrm{Im}[p_{2j}]\cos^{2j}(\theta). \tag{S13}$$

This is of the stated form. $\qquad\square$

Since the transformation from an operator to its canonical profile is linear, we have the following corollary:

**Corollary S2** (Linear combinations of QSP operators of even length exhibit canonical expansion). *If an operator $A$ can be decomposed into*

$$A = \sum_i \alpha_i U_i, \tag{S14}$$

*for $\alpha_i \in \mathbb{R}$ and QSP unitaries $U_i$ of even length (i.e. $A$ can be written as a linear combination of QSP unitaries of even length), then it admits a canonical expansion.*

**Remark S1.3** (Vector representation of canonical expansion). It will often be convenient to represent a canonical profile $\mathcal{P}$ in vector form. Assuming $\mathcal{P}_j^{(k)} = 0$ for all $j \geq d$, we can write the entire canonical profile as a vector in $\mathbb{R}^{4(d+1)}$,

$$\vec{\mathcal{P}}^{(k)} \equiv \begin{pmatrix} \vec{\mathcal{P}}_{d-1}^{(k)} \\ \vdots \\ \vec{\mathcal{P}}_0^{(k)} \\ \vec{\mathcal{P}}_{-1}^{(k)} \end{pmatrix}, \tag{S15}$$

where the vector $\vec{\mathcal{P}}_j^{(k)} \equiv (\mathcal{P}^{(0,k)}, \mathcal{P}_j^{(x,k)}, \mathcal{P}_j^{(y,k)}, \mathcal{P}_j^{(z,k)})$.

**Remark S1.4** (Canonical expansion of $Z$-rotation). Let $U_\epsilon$ be an operator admitting canonical expansion with vector

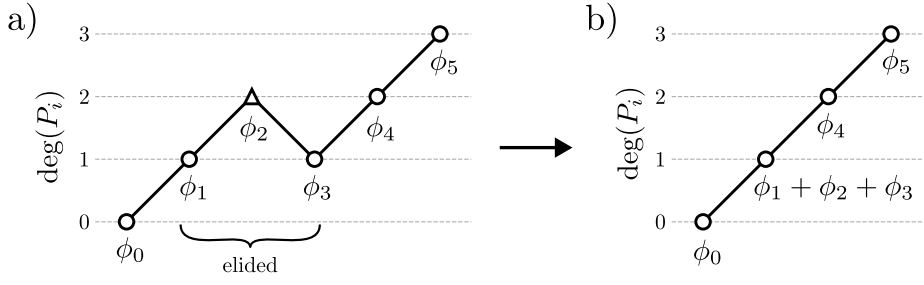

FIG. S1. The visualization of a) a length-5 QSP sequence parameterized by $(\phi_0, \cdots, \phi_5)$ and b) its length-3 elided form by the result of Lemma S4. We will refer to such plots in general as QSP degree plots.

form $\vec{\mathcal{P}}^{(k)}$ at order $k \geq 0$ (Remark S1.3). Then $V_\epsilon = e^{i\chi_0 Z} U_\epsilon e^{i\chi_1 Z}$ for $\chi_0, \chi_1 \in \mathbb{R}$ has canonical profile at order $k$

$$\vec{\mathcal{P}}^{(k)\prime} = O_z(\chi_0, \chi_1)\vec{\mathcal{P}}^{(k)} \equiv \begin{pmatrix} O_z(\chi_0, \chi_1) & & & & \\ & O_z(\chi_0, \chi_1) & & & \\ & & O_z(\chi_0, \chi_1) & & \\ & & & \ddots & \\ & & & & O_z(\chi_0, \chi_1) \end{pmatrix} \begin{pmatrix} \vec{\mathcal{P}}^{(k)}_{d-1} \\ \vdots \\ \vec{\mathcal{P}}^{(k)}_0 \\ \vec{\mathcal{P}}^{(k)}_{-1} \end{pmatrix}, \tag{S16}$$

where

$$O_z(\chi_0, \chi_1) \equiv \begin{pmatrix} \cos(\chi_0 + \chi_1) & 0 & 0 & -\sin(\chi_0 + \chi_1) \\ 0 & \cos(\chi_0 - \chi_1) & \sin(\chi_0 - \chi_1) & 0 \\ 0 & -\sin(\chi_0 - \chi_1) & \cos(\chi_0 - \chi_1) & 0 \\ \sin(\chi_0 + \chi_1) & 0 & 0 & \cos(\chi_0 + \chi_1) \end{pmatrix}. \tag{S17}$$

**Definition S1.5** (Degree-0 operator). We call a QSP unitary $U_\epsilon$ *degree-0 to order* $k \geq 1$ if it can be written

$$U_\epsilon = e^{i(z_0 + z_1\epsilon + O(\epsilon^2))Z + i\epsilon^k[(x+O(\epsilon))X + (y+O(\epsilon))Y]}, \tag{S18}$$

for some real $x$, $y$, $z_1$ independent of $\epsilon$ but possibly functions of $\theta$, and $z_0 \in \mathbb{R}$. Additionally, we call any QSP operator satisfying Eq. (S18) for some $k \geq 1$ *degree-0*. Equivalently, a QSP operator $U_\epsilon$ is degree-0 if $U_0 = e^{iz_0 Z}$ for some $z_0 \in \mathbb{R}$.

**Definition S1.6** (Unbiased operator). We call a QSP unitary $U_\epsilon$ *unbiased to order* $k \geq 1$ if it is degree-0 and $U_0 = I$.

**Remark S1.7** (Equivalence of expansions of unbiased operators to leading-order). Two different expansions in $\epsilon$ have been presented: the expansion in Definition S1.6 is performed in the exponent (i.e. the expansion is in the Hermitian generator of the unitary operator), while the canonical expansion (Definition S1.2) is of the unitary operator itself. These expansions do not yield the same expansion coefficients in general; however, for unbiased operators, the coefficients are identical to the leading-order for $k \geq 1$:

$$e^{i\epsilon(z+O(\epsilon))Z + i\epsilon^k[(x+O(\epsilon))X + (y+O(\epsilon))Y]} = (1 + O(\epsilon^2))I + i\epsilon(z + O(\epsilon))Z + i\epsilon^k[(x + O(\epsilon))X + (y + O(\epsilon))Y]. \tag{S19}$$

As all of our analysis will be done recursively on the leading order, we will use these forms interchangeably.

### B. Diagrammatic notation and manipulations

In this Appendix, we introduce a diagrammatic notation that will be useful for visualizing and analyzing QSP error operators. To this end, we prove a number of lemmas that imply useful diagrammatic manipulations.

#### 1. Basic properties

**Lemma S3.** Let $U_0 = [\![P, Q]\!]$ be a length-$d$ QSP unitary and $k = \deg(P) \leq d$. The unitary $U_0' = U_0 e^{i\phi_0 Z} W e^{i\phi_1 Z} = [\![P', Q']\!]$ is a length-$(d+1)$ QSP unitary with either $\deg(P') = k - 1$ or $\deg(P') = k + 1$.

*Proof.* Computing the product, we find

$$P'(a) = e^{-i(\phi_0 + \phi_1)} \left( aP(a) - \left(1 - a^2\right) Q(a) e^{2i\phi_0} \right).$$ (S20)

Since $\deg(P) = k$ by assumption and $\deg(Q) = k - 1$, it must be that $\deg(P') \leq k + 1$.

Next, we prove that $\deg(P') \geq k - 1$ by contradiction. Assume that $\deg(P') = w < k - 1$. We can then iterate our construction above choosing $U_0'' = U_0' e^{-i(\phi_1 - \frac{\pi}{2})Z} W e^{-i(\phi_0 - \frac{\pi}{2})Z}$ and, by the above argument, we have $\deg(P'') \leq w + 1 < k$. But we have chosen the additional phases such that $U_0'' = U_0 I_0$ where $I_0$ is a irreducible unbiased QSP corresponding to $(\phi_0, \pi/2, -\phi_0 + \pi/2)$. Therefore, $\deg(P'') = k$. This is a contradiction and so it must be that $\deg(P') \geq k - 1$.

Furthermore, we have that $\deg(P') \neq k$ by parity constraints. Therefore, the Lemma follows. $\qquad\square$

### 2. The conjugation super-operator

**Definition S1.8** (The conjugation super-operator). Given $\eta \in \mathbb{R}$ and $m, n \in \mathbb{Z}$, we use $\mathcal{C}_{m,n,\eta}$ to denote the super-operator that maps a length-$d$ sequence $\mathrm{QSP}(\theta; \vec{\phi})$ to

$$\mathcal{C}_{m,n,\eta} \mathrm{QSP}(\theta; \vec{\phi}) \equiv e^{-i(\eta + \pi(2m+n+\frac{1}{2}))Z} W e^{i\pi(n+\frac{1}{2})Z} \mathrm{QSP}(\theta; \vec{\phi}) W e^{i\eta Z},$$ (S21)

which is a length-$(d + 2)$ QSP sequence with phase angles $-(\eta + \pi(2m + n + \frac{1}{2}))$, $\pi(n + \frac{1}{2}) + \phi_0, \phi_1, \ldots, \phi_d$, and $\eta$.

The conjugation $\mathcal{C}_{m,n,\eta}$ super-operator appears naturally in our analysis of the error operator and subsequent construction of the recovery sequence.

We note several useful properties of the conjugation operation in the following remarks:

**Remark S1.9** (Recurrence under conjugation, first-order). Given an unbiased operator $U_\epsilon$ with functions $w, x, y, z$ in its canonical expansion, the corresponding functions of conjugated operator $U_\epsilon' = \mathcal{C}_{m,n,\eta} U_\epsilon$ are given by

$$\begin{pmatrix} w' \\ x' \\ y' \\ z' \end{pmatrix} = \begin{pmatrix} \cos^2(\theta) & 0 & 0 & 0 \\ 0 & \cos(2\eta) & -\cos^2(\theta)\sin(2\eta) & \cos^2(\theta)\sin(2\eta) \\ 0 & \sin(2\eta) & \cos(2\theta)\cos(2\eta) & -\cos^2(\theta)\cos(2\eta) \\ 0 & 0 & 4\sin^2(\theta) & \cos(2\theta) \end{pmatrix} \begin{pmatrix} w \\ x \\ y \\ z \end{pmatrix}.$$ (S22)

From Eq. (S22), we can write the recurrence of the canonical profiles using the vector notation of Remark S1.3. Suppose there exists some $d \geq 0$ such that the canonical profile of $U_\epsilon$ satisfies $\mathcal{P}_j^{(\sigma,1)} = 0$ for all $j \geq d$, then recurrence of the canonical profile of $U_\epsilon'$ satisfies

$$\begin{pmatrix} \vec{\mathcal{P}}_d'^{(1)} \\ \vec{\mathcal{P}}_{d-1}'^{(1)} \\ \vdots \\ \vec{\mathcal{P}}_0'^{(1)} \\ \vec{\mathcal{P}}_{-1}'^{(1)} \end{pmatrix} = \begin{pmatrix} A(\eta) \\ B(\eta) & A(\eta) \\ & B(\eta) & A(\eta) \\ & & & \ddots \\ & & & B(\eta) & A(\eta) \\ & & & & B(\eta) \end{pmatrix} \begin{pmatrix} \vec{\mathcal{P}}_{d-1}^{(1)} \\ \vdots \\ \vec{\mathcal{P}}_0^{(1)} \\ \vec{\mathcal{P}}_{-1}^{(1)} \end{pmatrix}$$ (S23)

where the matrix on the right-hand side is a block bidiagonal matrix of size $4(d + 2) \times 4(d + 1)$ with blocks

$$A(\eta) \equiv \begin{pmatrix} 0 & 0 & 0 & 0 \\ 0 & 0 & -2\sin 2\eta & \sin 2\eta \\ 0 & 0 & 2\cos 2\eta & \cos 2\eta \\ 0 & 0 & -4 & 2 \end{pmatrix}, \quad B(\eta) \equiv \begin{pmatrix} 1 & 0 & 0 & 0 \\ 0 & \cos 2\eta & \sin 2\eta & 0 \\ 0 & \sin 2\eta & -\cos 2\eta & 0 \\ 0 & 0 & 4 & -1 \end{pmatrix}.$$ (S24)

**Remark S1.10** (Linearity of conjugation). Note that conjugation is linear. For operators $U$, $V$ and coefficients $\alpha, \beta \in \mathbb{R}$,

$$\mathcal{C}_{m,n,\eta}(\alpha U + \beta V) = \alpha \mathcal{C}_{m,n,\eta} U + \beta \mathcal{C}_{m,n,\eta} V.$$ (S25)

### 3. The elision operation

**Remark S1.11** (QSP sub-sequence notation). For QSP $U$ of length-$d$ parameterized by $\vec{\phi} \equiv (\phi_0, \ldots, \phi_d)$, we will use notation $\vec{\phi}_{i:j}$ to denote the sub-sequence $(\phi_i, \ldots, \phi_j)$, and $U^{(j)}$ to denote the length-$j$ QSP parameterized by $\vec{\phi}_{0:j}$.

**Definition S1.12** (QSP degree peak). Let $R$ be an unbiased QSP sequence of length $d \geq 2$ parameterized by $(\phi_0, \ldots, \phi_d) \in \mathbb{R}^{d+1}$. Suppose for some $0 < i < d$ we have $\deg(P_{R^{(i)}}) = r + 1$ and $\deg(P_{R^{(i-1)}}) = \deg(P_{R^{(i+1)}}) = r$. We will call position $i$ a *degree peak* of $R$.

**Lemma S4** (QSP elision). *Let $R$ be a QSP operator of length $d \geq 2$ parameterized by $(\phi_0, \ldots, \phi_d)$. Position $i$ is a degree peak of $R$ if and only if $\phi_i = \pi\left(n + \frac{1}{2}\right)$ for some $n \in \mathbb{Z}$.*

*Additionally, $R$ is equivalent to a length-$(d-2)$ QSP parameterized by phases*

$$(\phi_0, \ldots, \phi_{i-2}, \phi_{i-1} + \phi_i + \phi_{i+1}, \phi_{i+2}, \ldots, \phi_d). \tag{S26}$$

*We refer to this transformation as* QSP elision.

*Proof.* Writing out the product, we find the following relationship between QSP polynomials of $R^{(i-1)}$ and $R^{(i)}$:

$$P_{R^{(i)}} = ae^{i(\phi_{i-1} + \phi_i)}(P_{R^{(i-1)}} + e^{-2i\phi_{i-1}}Q_{R^{(i-1)}}) + \Theta(a^{r-1}). \tag{S27}$$

By assumption $\deg(P_{R^{(i)}}) = r + 1$ and therefore $\deg(P_{R^{(i-1)}} + e^{-2i\phi_{i-1}}Q_{R^{(i-1)}}) = \deg(P_{R^{(i)}}) - 1 = r$.

Writing out the product for $R^{(i+1)}$, we find

$$P_{R^{(i+1)}} = 2a^2 \cos(\phi_i) e^{i(\phi_{i-1} + \phi_{i+1})}(P_{R^{(i-1)}} + e^{-2i\phi_{i-1}}Q_{R^{(i-1)}}) + \Theta(a^r). \tag{S28}$$

Looking at the equation above and comparing with the $P_{R^{(i)}}$ we find that if $\deg(P_{R^{(i+1)}}) = r$, it must be that $\cos(\phi_i) = 0$ or equivalently $\phi_i = \pi\left(n + \frac{1}{2}\right)$ for some $n \in \mathbb{Z}$. The converse is also true.

Assuming $\phi_i = \pi(n + \frac{1}{2})$, we find,

$$P_{R^{(i+1)}} = e^{i(\phi_{i-1} + \phi_i + \phi_{i+1})}P_{R^{(i-1)}}, \tag{S29}$$

This transformation is equivalent to a $Z$-rotation by $\phi_{i-1} + \phi_i + \phi_{i+1}$ (a length-0 QSP). We can therefore elide the original QSP sequence by replacing the three phases $\phi_{i-1}$, $\phi_i$, and $\phi_{i+1}$ with a single phase $\phi_{i-1} + \phi_i + \phi_{i+1}$, thus proving the Lemma. $\qquad\square$

We then have the following as a corollary:

**Corollary S5** (Degree-0 length-2 QSPs). *A length-2 sequence $\mathrm{QSP}(\theta; (\phi_0, \phi_1, \phi_2))$ is degree-0 if and only if*

$$\phi_0 = \chi + \left(\phi + \pi\left(2m + n + \frac{1}{2}\right)\right), \tag{S30}$$

$$\phi_1 = \pi\left(n + \frac{1}{2}\right), \tag{S31}$$

$$\phi_2 = \phi, \tag{S32}$$

*for some $\phi, \chi \in \mathbb{R}$ and $n, m \in \mathbb{Z}$.*

*Equivalently,*

$$\mathrm{QSP}(\theta; (\phi_0, \phi_1, \phi_2)) = e^{i\chi Z}\mathcal{C}_{m,n,\phi}I. \tag{S33}$$

The construction of inverse QSP operators can also be seen as a immediate consequence of Lemma S4.

**Corollary S6** (Inverse QSPs). *Let $U$ be length-$d$ QSP operator parameterized by phases $(\phi_0, \ldots, \phi_d)$. The length-$d$ QSP $U'$ parameterized by $(-\phi_d + \frac{\pi}{2}, -\phi_{d-1}, \ldots, -\phi_1, -\phi_0 - \frac{\pi}{2})$ is the inverse QSP sequence in the sense that $UU' = U'U = I$.*

**Remark S1.13** (Summary of diagrammatic notation). We can summarize the results of this section using a concise diagrammatic notation. An example of such a plot is given in Fig. S1 and has several notable features:

- Arbitrary $Z$-rotations are represented by open circles, and we use open triangles to plot QSP phases that are a half-integer multiple of $\pi$ to distinguish degree-peaks.

- Markers with solid fill are used to indicate additional rotations by $\pi/2$. Markers with checkerboard fill are used to indicate $\epsilon$-noisy rotations.

- Signal operators are represented by solid lines.

- The vertical axis is used to plot the degree of the polynomial $P_{U^{(i)}}$ at position $i$;

- By Lemma S3, each layer of signal processing either increases or decreases the degree of the polynomial $P_i(a) = \langle 0 | U(\theta, \vec{\phi}_{0;i}) | 0 \rangle$ by exactly one.

- Finally, Lemma S4 provides us with a way of simplifying QSP diagrams with degree-peaks through elision.

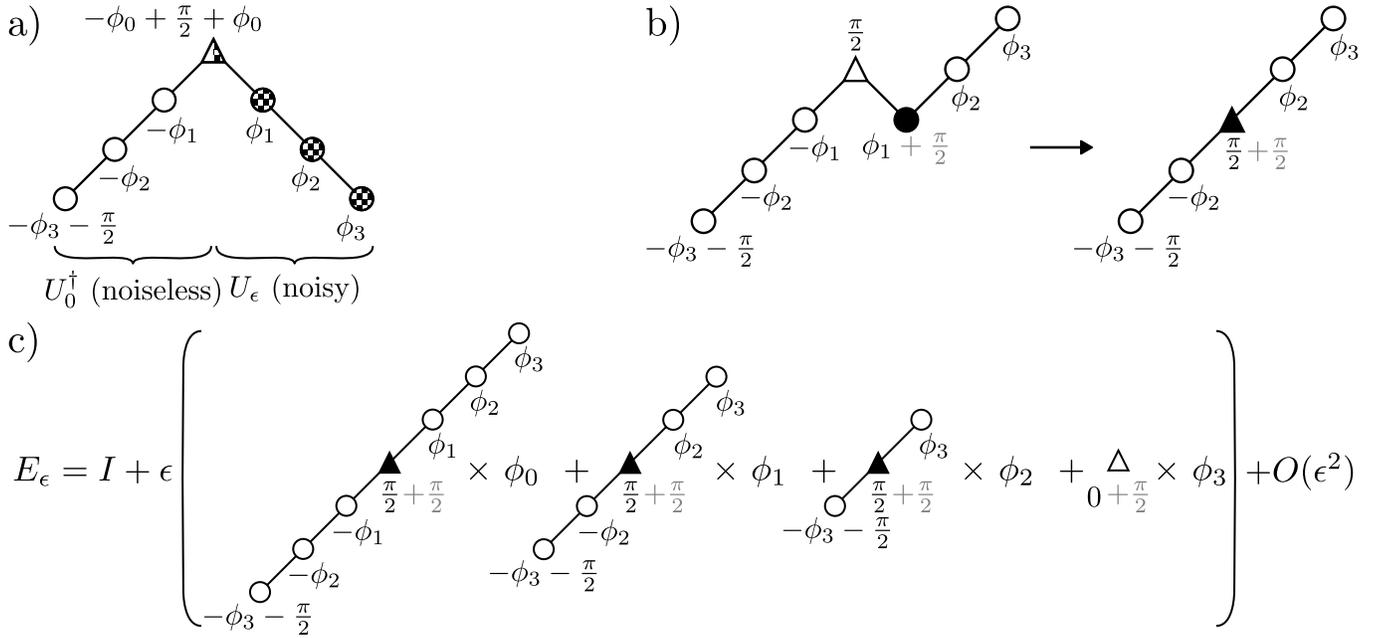

FIG. S2. Diagrammatic representation of an error operator for a length-3 QSP parameterized by $\phi_i \in \mathbb{R}$. a) Error operator $E_\epsilon = U_0^\dagger U_\epsilon$. Checkerboard fill indicates $\epsilon$-noisy rotations (note the peak phase is only partially noisy). b) Analysis of one term in the first-order perturbative expansion of the error operator corresponding to an over-rotation error of the $\phi_1$ phase and elided form. The location of $\frac{\pi}{2}$ over-rotation errors is marked by filled markers. c) Expansion of the error operator showing all diagrams to first-order with corresponding weights.

## S2. Analysis of degree-0 operators

In this Appendix we analyze two notable types of degree-0 operator: firstly, the error operator of an arbitrary QSP (Remark S2.2); and the noisy degree-0 QSP (Remark S2.8), which will become the basis of our recovery construction. To this end, we make use of the diagrammatic techniques developed in Section S1.

### A. Analysis of error operators

Given a noisy QSP, we would like to isolate the error term. We accomplish this by pre-multiplying the noisy QSP operator by its noiseless inverse. By definition, the error must be a degree-0 operator as it vanishes as $\epsilon \to 0$ (Definition S1.5).

**Definition S2.1** (Error operator). Given a QSP operator $U$, we define its *error operator* $E_\epsilon = U_0^\dagger U_\epsilon$. The inverse operator $U_0^\dagger$ can be parameterized as a QSP operation using Corollary S6. Note that the error operator is an unbiased operator as $E_0 = I$.

We now perform a perturbative analysis of the error operators of QSPs under our noise model.

**Remark S2.2** (Perturbative expansion of the error operator). We can expand a noisy $Z$-rotation in orders of $\epsilon$ as follows:

$$e^{i\phi(1+\epsilon)Z} = e^{i\phi Z} \prod_{k=0}^{\infty} \epsilon^k \phi^k e^{i\frac{\pi k}{2}Z}. \tag{S34}$$

Substituting into Eq. (S1), we obtain

$$U_\epsilon(\theta; \vec{\phi}) = \left( e^{i\phi_0 Z} \prod_{k_0=0}^{\infty} \epsilon^{k_0} \phi^{k_0} e^{i\frac{\pi}{2}k_0 Z} \right) \prod_{j=1}^{d} \left[ W(\theta) \left( e^{i\phi_j Z} \prod_{k_j=0}^{\infty} \epsilon^{k_j} \phi^{k_j} e^{i\frac{\pi}{2}k_j Z} \right) \right]. \tag{S35}$$

Rewriting in orders of $\epsilon$,

$$
\begin{aligned}
U_\epsilon(\theta; \vec{\phi}) = U_0 + \epsilon \Big( &\phi_0 e^{i(\phi_0 + \frac{\pi}{2})Z} W e^{i\phi_1 Z} W \ldots W e^{i\phi_d Z} \\
&+ \phi_1 e^{i\phi_0 Z} W e^{i(\phi_1 + \frac{\pi}{2})Z} W \ldots W e^{i\phi_d Z} \\
&+ \ldots \\
&+ \phi_d e^{i\phi_0 Z} W e^{i\phi_1 Z} W \ldots W e^{i(\phi_d + \frac{\pi}{2})Z} \Big) + O(\epsilon^2).
\end{aligned}
\tag{S36}
$$

The first-order term is a sum of $d+1$ QSP unitaries, each a copy of the noiseless QSP with a $\frac{\pi}{2}$ over-rotation at location $j$ weighted by $\phi_j$ for each index $j$. Likewise, the $k^{\text{th}}$-order term is a sum of $(d+1)^k$ QSP unitaries corresponding to all possible ways to to insert $k$ over-rotations by $\frac{\pi}{2}$ (including multiple over-rotations at the same index).

To obtain the error operator of Definition S2.1, we left-multiply by $U_0^\dagger$ which can be written as a noiseless QSP per the construction of Corollary S6. Each of these sequences can be simplified using repeated application of Lemma S4. To first-order, the result is

$$
\begin{aligned}
E_\epsilon = I + \epsilon \big( &\phi_0 \times \mathrm{QSP}(\theta; (-\phi_d - \pi/2, -\phi_{d-1}, \ldots, -\phi_2, -\phi_1, \pi, \phi_1, \phi_2, \ldots, \phi_{d-1}, \phi_d)) \\
&+ \phi_1 \times \mathrm{QSP}(\theta; (-\phi_d - \pi/2, -\phi_{d-1}, \ldots, -\phi_2, \pi, \phi_2, \ldots, \phi_{d-1}, \phi_d)) \\
&+ \ldots \\
&+ \phi_{d-1} \times \mathrm{QSP}(\theta; (-\phi_d - \pi/2, \pi, \phi_d)) \\
&+ \phi_d \times \mathrm{QSP}(\theta; (\pi/2)) \big) + O(\epsilon^2).
\end{aligned}
\tag{S37}
$$

Note that the first-order expansion in Eq. (S37) consists of a weighted sum of even length QSP unitaries of a special form, which we generalize in Definition S2.3. Analogous calculations show that, the higher-order terms in the expansion are likewise weighted sums over QSP unitaries of even length. Therefore by Corollary S2, the error operator $E_\epsilon$ admits a canonical expansion.

An example in diagrammatic form is provided for a general length-3 QSP in Fig. S2.

We generalize the form of the first-order error found in Eq. (S37) with the following:

**Definition S2.3** (Standard form, first-order). We say an operator $A$ is in first-order *standard form* of degree-$2d$ if it is written as a weighted sum over QSP operators generated by conjugation of $e^{i\frac{\pi}{2}Z}$

$$
\begin{aligned}
A = &\alpha_0 \times \mathrm{QSP}(\theta; (-\phi_d - \pi/2, -\phi_{d-1}, \ldots, -\phi_2, -\phi_1, \pi, \phi_1, \phi_2, \ldots, \phi_{d-1}, \phi_d)) \\
&+ \alpha_1 \times \mathrm{QSP}(\theta; (-\phi_d - \pi/2, -\phi_{d-1}, \ldots, -\phi_2, \pi, \phi_2, \ldots, \phi_{d-1}, \phi_d)) \\
&+ \ldots \\
&+ \alpha_{d-1} \times \mathrm{QSP}(\theta; (-\phi_d - \pi/2, \pi, \phi_d)) \\
&+ \alpha_d \times \mathrm{QSP}(\theta; (\pi/2)).
\end{aligned}
\tag{S38}
$$

where $\alpha_i \in \mathbb{R}$ and $\phi_i \in \mathbb{R}$.

**Definition S2.4** (Error profile). Let $U_\epsilon$ be a noisy operator, we will refer to the canonical profile of its error operator as the *error profile* of $U_\epsilon$. Note that the analysis of Remark S2.2 shows that if $U_\epsilon$ is a noisy QSP operator, it admits an error profile of the form in Definition S1.2.

**Remark S2.5** (Canonical profile of product of unbiased operators, leading-order). If $U_\epsilon, V_\epsilon$ are both unbiased to order $k \geq 1$ with canonical profiles $\mathcal{P}$ and $\mathcal{Q}$ respectively, then $W_\epsilon = U_\epsilon V_\epsilon$ is also unbiased to order $k$ with an canonical profile $\mathcal{R}$ satisfying

$$
\mathcal{R}_j^{(\sigma, k)} = \mathcal{P}_j^{(\sigma, k)} + \mathcal{Q}_j^{(\sigma, k)}, \qquad \sigma \in \{x, y\},
\tag{S39}
$$

$$
\mathcal{R}_j^{(\sigma, 1)} = \mathcal{P}_j^{(\sigma, 1)} + \mathcal{Q}_j^{(\sigma, 1)}, \qquad \sigma \in \{0, z\},
\tag{S40}
$$

$$
\tag{S41}
$$

for all $j$.

**Remark S2.6** (Error profile of degree-0 operators). A degree-0 QSP operator $R_\epsilon$ with $R_0 = e^{i\chi Z}$ for $\chi \in \mathbb{R}$ has error operator $E_\epsilon = e^{-i\chi Z} R_\epsilon$. That is, a degree-0 QSP operator's error profile is a rotated version of its own canonical profile, which can be calculated using Remark S1.4.

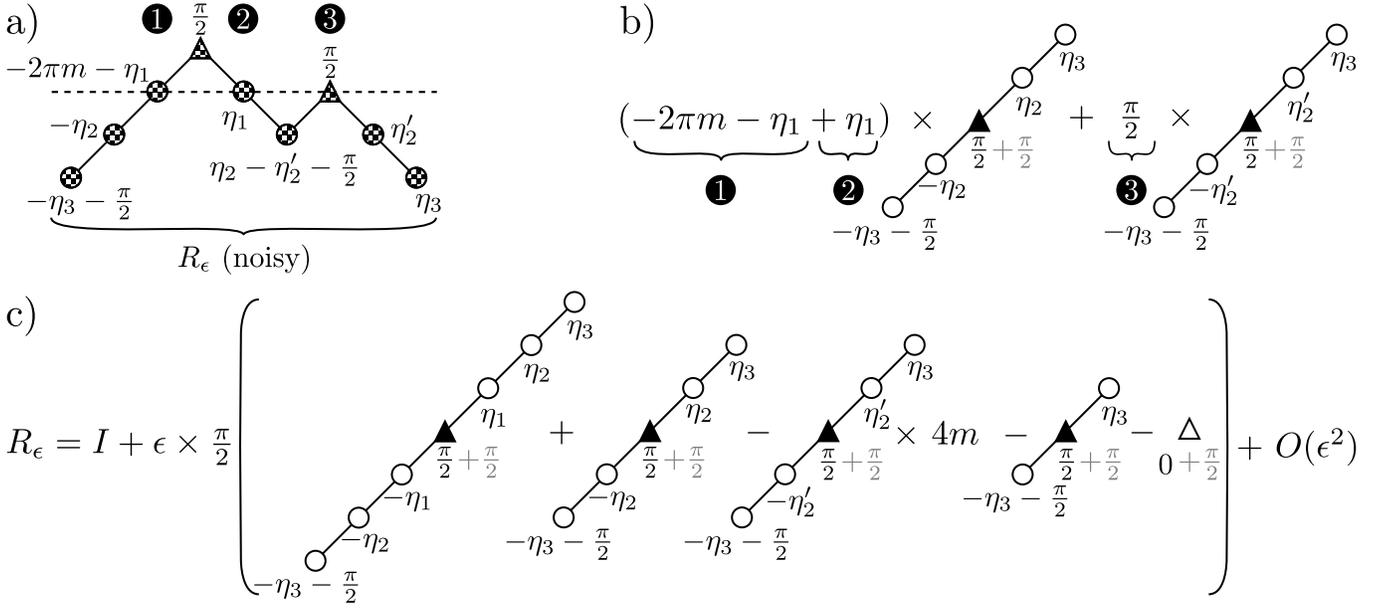

FIG. S3. Diagrammatic analysis of an irreducible degree-0 QSP parameterized by $\chi = 0$, $\eta_i \in \mathbb{R}$ and $m \in \mathbb{Z}$ (compare with Fig. S2). a) Diagrammatic representation of an irreducible degree-0 QSP $R_\epsilon$. A detailed analysis is performed for over-rotation errors occurring at select locations (labels 1–3) corresponding to errors at degree-2 (dashed line). Checkerboard fill indicates $\epsilon$-noisy rotations. b) Analysis of diagrams resulting from over-rotation errors at locations labeled in previous sub-figure after elision. The location of $\frac{\pi}{2}$ over-rotation errors is marked by filled markers. c) Expansion of the recovery operator showing all diagrams to first-order with corresponding weights. Notice that weights are integer multiples of $\pi/2$.

## B. Analysis of recovery operators

In this subsection we will study the basic components of our error recovery sequence. To this end, we study the properties of degree-0 QSP operators, and in particular, the properties of their irreducible building blocks.

**Definition S2.7.** A degree-0 QSP unitary of length $d$ is called *irreducible* if $\deg(P^{(i)}) > 0$ for all $0 < i < d$; otherwise an degree-0 QSP is called *reducible*.

**Lemma S7** (Decomposition of irreducible degree-0 QSP unitary). *An irreducible degree-0 QSP sequence $R$ of length $d \geq 2$ parameterized by phases $\vec{\phi} \in \mathbb{R}^{d+1}$ can be written as $R = e^{i\chi Z}\mathcal{C}_{m,n,\phi_d}R'$ for some $\chi \in \mathbb{R}$ and $m, n \in \mathbb{Z}$, and unbiased QSP $R'$ of length-$(d-2)$.*

*Proof.* If $d = 2$, then $R = e^{i\chi Z}\mathcal{C}_{m,n,\phi_d}I$ for some $\chi \in \mathbb{R}$ and $m, n \in \mathbb{Z}$ by Corollary S5.

For $d > 2$, we proceed by repeated application of the QSP elision operation (Lemma S4), each time reducing the length of $R$ by 2. In particular, since the unbiased QSP $R$ is irreducible, it has a degree peak at location $2 \leq i \leq d-2$. Performing elision about position $i$, we are left with the length-$(d-2)$ irreducible degree-0 QSP sequence. Notably, neither phases $\phi_0$ nor $\phi_d$ are affected by performing elision at location $2 \leq i \leq d-2$. After $(d/2-1)$ elision steps, we are left with a length-2 QSP parameterized by $\text{QSP}(\theta; (\phi_0, \sum_{i=1}^{d-1}\phi_i, \phi_d))$, where by Lemma S4, we have that

$$\sum_{i=1}^{d-1}\phi_i = \pi\left(n + \frac{1}{2}\right), \tag{S42}$$

for some $n \in \mathbb{Z}$.

Therefore $\text{QSP}(\theta, \vec{\phi}_{1:d-1}) = e^{i\pi(n+\frac{1}{2})Z}$ or equivalently $e^{-i\pi(n+\frac{1}{2})Z}\text{QSP}(\theta, \vec{\phi}_{1:d-1}) = I$. Thus we can rewrite the original QSP in the desired form $R = e^{i\chi Z}\mathcal{C}_{m,n,\phi_d}R'$ for $\chi \in \mathbb{R}$ and unbiased length-$(d-2)$ QSP $R' \equiv \text{QSP}(\theta; (\phi_1 - \pi(n + \frac{1}{2}), \phi_2, \ldots, \phi_{d-1}))$ proving the Lemma. $\square$

**Remark S2.8** (Expansion of a degree-0 QSP, first-order). As noted in Remark S2.6, an degree-0 operator is a rotated version of its own error operator. Therefore $R_\epsilon$ can be written a form similar to that of Definition S2.3. We aim to show that for the case of degree-0 $R_\epsilon$, these weights are additionally integer multiples of $\frac{\pi}{2}$ save for the degree-0 term.

First, consider the case of irreducible degree-0 QSP $R = \text{QSP}(\theta; (\phi_0, \ldots, \phi_d))$. Consider the contributions to the first-order error from over-rotations at the first and last positions (i.e. assume for now errors do not affect positions $0 < i < d$). By Lemma S7 we can write irreducible

$$R = e^{i\chi Z} \mathcal{C}_{m,n,\phi_d} R' = e^{i\phi_0 Z} W e^{i\pi(n+\frac{1}{2})Z} R' W e^{i\phi_d Z}, \tag{S43}$$

for $R'$ unbiased $\chi \in \mathbb{R}$ and $m, n \in \mathbb{Z}$. Over-rotation at the first and last positions occur at degree-0 and therefore both produce an degree-0 error term equivalent to $e^{i(\chi+\frac{\pi}{2})Z}$ and the overall weight of the degree-0 diagram is $\phi_0 + \phi_d = \chi - \frac{\pi}{2}(4m + 2n + 1)$. The same analysis holds for the unbiased $R'$, however $\phi_i + \phi_{d-i}$ must be an integer multiple of $\pi/2$ for $0 < i < d$ by Lemma S4, and therefore the error diagram must have weight that is a integer multiple of $\pi/2$. Furthermore, the weights are preserved by the linearity of error profile to first-order under conjugation (Remark S1.10) and product (Remark S2.5). Thus, the same holds for higher degree error diagrams.

Now consider a general degree-0 QSP $R$ decomposed into $r$ constituent irreducible components,

$$R = e^{i\chi_1 Z} J^{(1)} e^{i\chi_2 Z} J^{(2)} \ldots e^{i\chi_r Z} J^{(r)}, \tag{S44}$$

where $\chi_1, \ldots, \chi_r \in \mathbb{R}$, and each $J^{(j)}$ is an irreducible unbiased QSP. We can in general write an degree-0 QSP $R_\epsilon$ up to first-order in $\epsilon$ as follows:

$$R_\epsilon = e^{i\chi Z} \left[ I + \left( \sum_i c_i e^{i\frac{\pi}{2}Z} + \frac{\pi}{2} \sum_i d_i U_i \right) + O(\epsilon^2) \right], \tag{S45}$$

for $\chi = \chi_1 + \cdots + \chi_r$, $c_i \in \mathbb{R}$, $d_i \in \mathbb{Z}$ and QSP unitaries $U_i$ of even length. Additionally, each $U_i$ is of the form

$$U_i = \mathcal{C}_{m_{i,d}, n_{i,d}, \eta_{i,d}} \ldots \mathcal{C}_{m_{i,1}, n_{i,1}, \eta_{i,1}} e^{i\frac{\pi}{2}}. \tag{S46}$$

A diagrammatic analysis is provided for an example length-8 irreducible degree-0 QSP in Fig. S3.

## S3.  Proof of Theorem 1: Z-error is not correctable in general

In this Appendix, we prove the impossibility of the most general form of error correction (Theorem 1), thereby motivating our restricted definition, which we dub '$XY$ error correction' (Theorem 3).

**Remark S3.1** (Necessary condition for general QSP recovery). Let $U_\epsilon$ be a noisy QSP unitary. The most general form of error correction would be to seek another unitary $U'_\epsilon$, itself a noisy QSP operator, such that for some $k \geq 1$, for all states $|\psi\rangle$,

$$|\langle\psi|U'_\epsilon|\psi\rangle|^2 = |\langle\psi|U_0|\psi\rangle|^2 + O(\epsilon^{k+1}). \tag{S47}$$

This is equivalent to requiring

$$U'_\epsilon = U_0 e^{i\chi} e^{i\epsilon^{k+1}(xX+yY+zZ+O(\epsilon))}, \tag{S48}$$

for some global phase $\chi \in \mathbb{R}$ and $x$, $y$, and $z$ functions of $\theta$.

We will show that this is generally not possible in Theorem 1.

We continue with a few results needed in our proof of the impossibility result.

**Lemma S8** (Bottom-degree term of degree-0 QSP). *Let $U_\epsilon$ be a degree-0 QSP with $U_0 = e^{i\chi Z}$ for $\chi \in \mathbb{R}$ (as required by Remark S3.1). Then the bottom-degree $Z$ term in its error profile $\mathcal{P}$ to first-order in $\epsilon$ is*

$$\mathcal{P}^{(z,1)}_{-1} = (\chi + m\pi)\cos\chi, \tag{S49}$$

*for some $m \in \mathbb{Z}$.*

*Proof.* The expansion of a general degree-0 QSP $U$ to first-order is given by Eq. (S45). Remark S1.9 gives us the lowest degree $Z$ coefficients in the canonical expansion of each constituent diagram: the contribution to the lowest degree $Z$ term is a product of the bottom-left component of the $B$ matrices of Eq. (S24), which for all degree $\geq 2$ diagrams is 1 and for the degree-0 diagram $e^{i\frac{\pi}{2}Z}$ is $-1$. After a careful accounting of the weights, we find that the sum from all diagrams to this lowest degree term is $\chi + m\pi$ for $m \in \mathbb{Z}$. Finally, the overall $e^{i\chi X}$ rotation of the first-order term in Eq. (S45) results in an overall multiplicative factor of $\cos\chi$ on the expansion by Remark S1.4. Together, this results in $\mathcal{R}^{(z,1)}_{-1}$ of the form required by the Lemma. $\qquad\square$

We are now ready to prove Theorem 1 which is restated below.

**Theorem 1** (No correction of Z-error). *Let $U_\epsilon$ be a length-$d$ noisy QSP unitary parameterized by $(\phi_0, \ldots, \phi_d) \in \mathbb{R}^{d+1}$. For general phases $\phi_i$, no noisy QSP unitary $U'_\epsilon$ exists such that for any $k \geq 1$,*

$$\langle 0|U'_\epsilon|0\rangle = \langle 0|U_0|0\rangle + O(\epsilon^{k+1}). \tag{S50}$$

*Proof.* First we show that we cannot fully recover a noisy length-0 QSP $U_\epsilon \equiv e^{i\phi_0(1+\epsilon)Z}$ to first-order for general $\phi_0 \in \mathbb{R}$. For full recovery, the QSP must satisfy Eq. (S48), and therefore either $U'_0 = e^{i\phi_0 Z}$ or $U'_0 = e^{i(\phi_0+\pi)Z} = -e^{i\phi_0 Z}$; therefore it $U'$ must be a degree-0 QSP. We see immediately that its bottom-degree term, given by Lemma S8, cannot be corrected in general (i.e. unless $\phi_0$ is an integer multiple of $\pi/2$).

Now we generalize the result for QSPs of length $d > 0$. For contradiction, suppose that there exists an error correction function EC : $\mathbb{R}^{d+1} \to \mathbb{R}^{d'+1}$ for some $d' > d$ that is capable of mapping an arbitrary length-$d$ QSP sequence to one that is corrected to first-order. That is, suppose for any $\vec{\phi} \in \mathbb{R}^{d+1}$ parameterizing QSP $U_\epsilon = \text{QSP}_\epsilon(\theta, \vec{\phi})$ we have $\vec{\psi} = \text{EC}(\vec{\phi})$ and $U'_\epsilon = \text{QSP}_\epsilon(\theta, \vec{\psi})$ satisfying Eq. (S48). We can simulate a length-0 QSP operator $\text{QSP}(\theta, (\phi_0))$ by appending a recovered length-$d$ QSP and a recovered version of its inverse (Corollary S6). For concreteness, we can choose phases $\vec{\phi}_1, \vec{\phi}_2 \in \mathbb{R}^{d+1}$,

$$\vec{\phi}_1 = (-\pi/2, 0, \ldots, 0, \pi/2), \tag{S51}$$

$$\vec{\phi}_2 = (0, \ldots, 0, \phi_0). \tag{S52}$$

Let $R_\epsilon = \text{QSP}_\epsilon(\theta, \vec{\phi}_1)$ and $S_\epsilon = \text{QSP}_\epsilon(\theta, \vec{\phi}_2)$. Further let $\vec{\psi}_1 = \text{EC}(\vec{\phi}_1)$, $\vec{\psi}_2 = \text{EC}(\vec{\phi}_2)$, and $R'_\epsilon = \text{QSP}_\epsilon(\theta, \vec{\psi}_1)$ and $S'_\epsilon = \text{QSP}_\epsilon(\theta, \vec{\psi}_2)$. Note that by construction $R_0 S_0 = e^{i\phi_0 Z}$ as desired and therefore $R'_0 S'_0 = e^{i\phi_0 Z}$. Further, if both $R'_\epsilon$ and $S'_\epsilon$ satisfy Eq. (S48) for $k \geq 1$, then resulting length-$2d'$ QSP $R'_\epsilon S'_\epsilon$ will also be fully corrected to order $k$. This contradicts our original result for $d = 0$ and therefore EC cannot exist for any $d \geq 0$ thus proving the Theorem. $\qquad\square$

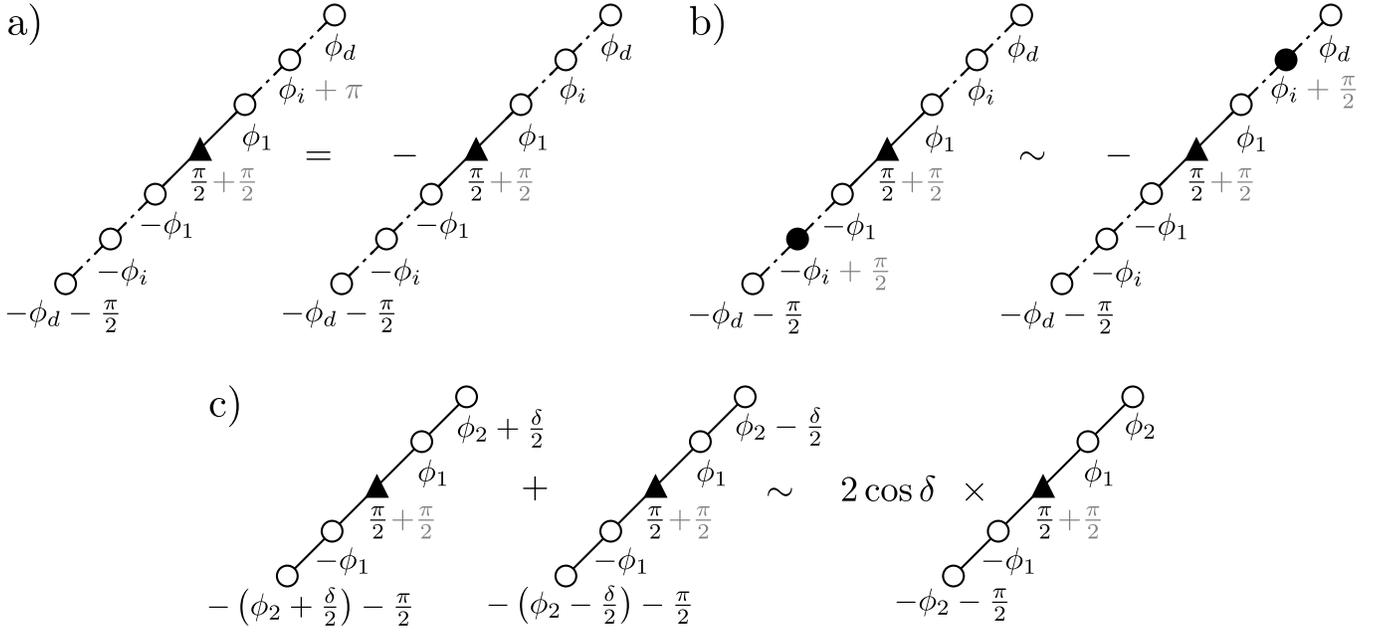

FIG. S4. Diagrammatic representations of diagrammatic equivalence relations for a) exact equivalence for over-rotation by $\pi$; b) $XY$-equivalence of conjugation by $\pi/2$; and c) $XY$-equivalence of counter-rotated diagrams.

## S4. Proof of Theorems 2 and 3: Higher-order Component-wise Recovery

In this Appendix, we generalize the first-order component-wise recovery construction of Section S4 A to all orders in $k$.

### A. First-order recovery

In light of the impossibility result presented in Section S3, we shift our attention to XY error recovery. In this Appendix, we show that it is possible to perform this restricted form of recovery and make use of the tools developed in Section S2 to provide a general construction for XY recovery operators.

**Remark S4.1** (Necessary condition for XY recovery). Let $U_\epsilon$ be a noisy QSP unitary. Our restricted error correction (i.e. XY recovery) condition is to seek another unitary $U'_\epsilon$, itself a noisy QSP operator, such that for some $k \geq 1$,

$$|\langle 0|U'_\epsilon|0\rangle|^2 = |\langle 0|U_0|0\rangle|^2 + O(\epsilon^{k+1}) \tag{S53}$$

This is equivalent to requiring

$$U'_\epsilon = U_0 e^{i(\chi+O(\epsilon))Z + \epsilon^{k+1}((x+O(\epsilon))X + (y+O(\epsilon))Y)} \tag{S54}$$

for some $\chi \in \mathbb{R}$, and $x$, $y$, and $z$ functions of $\theta$.

We make the following definition in light of Remark S4.1:

**Definition S4.2** (XY-equivalence). We say that two operators $U = w(\theta)I + i[x(\theta)X + y(\theta)Y + z(\theta)Z]$ and $V = w'(\theta)I + i[x'(\theta)X + y'(\theta)Y + z'(\theta)Z]$ are $XY$-equivalent if $x(\theta) = x'(\theta)$ and $y(\theta) = y'(\theta)$. We denote this $U \sim V$.

**Remark S4.3** (Counter-rotation identity). Here we make use of the following trigonometric identity

$$\begin{pmatrix} \sin(\eta+\delta) \\ -\cos(\eta+\delta) \end{pmatrix} + \begin{pmatrix} \sin(\eta-\delta) \\ -\cos(\eta-\delta) \end{pmatrix} = 2\cos(\delta)\begin{pmatrix} \sin(\eta) \\ -\cos(\eta). \end{pmatrix} \tag{S55}$$

Thus by duplicating the length-$2r$ sequence in Eq. (S58) and counter-rotating each copy by an amount $\delta/2$, we can

construct a sequence that is $XY$-equivalent to a re-scaled version of the original. this can be verified by Remark S1.4. This is represented diagrammatically in Fig. S4 c).

**Remark S4.4** (Constructing first-order recovery sequences)**.** Conjugation allows us to extend first-order unbiased operators. Given an unbiased QSP sequence $R$, the conjugation $R' = \mathcal{C}_{m,n,\eta}R$ is an irreducible unbiased sequence for all $\eta \in \mathbb{R}$ and $n, m \in \mathbb{Z}$.

In particular, we will use the following length-$2d$ sequence for recovery to first-order:

$$\mathcal{C}_{m_d,n,\eta_d} \ldots \mathcal{C}_{m_1,n,\eta_1} I, \tag{S56}$$

where $m_1, \ldots, m_d, n \in \mathbb{Z}$, and $\eta_1, \ldots, \eta_d \in \mathbb{R}$.

For our construction, we further restrict ourselves to an appended recovery sequence. Namely, given noisy QSP operator $U_\epsilon$, we seek a noisy recovery operation $R_\epsilon$ such that $U'_\epsilon \equiv U_\epsilon R_\epsilon$ satisfies Eq. (S54). Additionally, we require unbiased (i.e. $R_0 = I$). We perform recovery component-wise.

**Remark S4.5** (First-order component-wise recovery)**.** There is a striking similarity between the error operator expansion in Remark S2.2 and the general degree-0 in Remark S2.8 which are visualized in Fig. S2 and Fig. S3 respectively. We take advantage of this fact to construct our recovery operator.

Consider a noisy QSP $U_\epsilon = \mathrm{QSP}(\theta; (\phi_0, \ldots, \phi_d))$. As noted in Remark S2.2, its error operator $E_\epsilon$ to first-order can be decomposed into a sum of even-length QSPs from length $0, 2, \ldots, 2d$. The length-$2r$ diagram is in general is

$$\phi_{d-r} \times \mathrm{QSP}(\theta; (-\phi_d - \pi/2, \ldots, -\phi_{d-r+1}, \pi, \phi_{d-r+1}, \ldots, \phi_d)). \tag{S57}$$

The length-$2r$ recovery operator of the form in Remark S4.4 to first-order

$$\begin{aligned}
\mathcal{C}_{0,n,\phi_d + \pi/2}\mathcal{C}_{0,n,\phi_{d-1}} \ldots \mathcal{C}_{0,n,\phi_{d-r+1}} I = I + \epsilon \Big( &-\frac{n\pi}{2} \times \mathrm{QSP}(\theta; (\pi/2)) \\
&+ \frac{n\pi}{2} \times \mathrm{QSP}(\theta; (-\phi_d - \pi/2, \ldots, -\phi_{d-r+1}, \pi, \phi_{d-r+1}, \ldots, \phi_d)) \Big) \\
&+ O(\epsilon^2),
\end{aligned} \tag{S58}$$

for some $n \in \mathbb{Z}$ to be specified later.

Note the similarities between the degree-$2r$ terms in Eq. (S57) and Eq. (S58). The recovery QSP is parameterized by the last $r$ phases of the original QSP sequence. An additional $\pi/2$ shift added to the final phase $\phi_d \mapsto \phi_d + \pi$ to negate the $X$ and $Y$ components of the degree-$2r$ term at first-order: this can be confirmed by applying Remark S1.4.

The only remaining challenge is to match the $\phi_{d-r}$ weight of the degree-$2r$ term in Eq. (S57). Here, we make use of Remark S4.3 to continuously control the $X$ and $Y$ components of the recovery operator's error profile. To fully cancel the length-$2r$ diagram in Eq. (S57), we append two length-$2r$ recovery QSPs

$$\mathcal{C}_{0,n,\phi_d + \pi/2 \pm \delta}\mathcal{C}_{0,n,\phi_{d-1}} \ldots \mathcal{C}_{0,n,\phi_{d-r+1}} I, \tag{S59}$$

choosing $n \in \mathbb{Z}$ such that there is a solution to $\delta = \frac{1}{2}\cos^{-1}\left(\frac{\phi_{d-r}}{n\pi}\right)$.

In summary we have canceled the degree-$2r$ diagram using a length-$4r$ QSP operator. We repeat this for each diagram of length-$2r$ in the first-order expansion of the error operator for $r \in \{2, 4, \ldots, 2d\}$; the length-0 term contributes only to the $Z$ component of the error, which can be ignored. Overall, the recovery of each diagram takes a length $\Theta(r)$ QSP and we need to correct $\Theta(d)$ diagrams, resulting in a final recovery operator of length $\Theta(d^2)$.

### B. Standard form

First, we motivate and introduce new nomenclature useful in the analysis of the error and recovery operators to higher-order in $\epsilon$.

**Definition S4.6** (Error component)**.** Let $U$ be length-$d$ QSP operator parameterized by real phases $(\phi_0, \ldots, \phi_d)$. Then a length-$2r$ QSP $V$ with $1 \leq r \leq d$ is said to be an *error component of QSP $U$* if it can be written in the following form:

$$V = \mathrm{QSP}(\theta; (-\phi_d - \pi/2, \ldots, -\phi_{d-r+1}, \pi, \phi_{d-r+1} + b_{d-r+1}\pi/2, \ldots, \phi_d + b_d\pi/2)), \tag{S60}$$

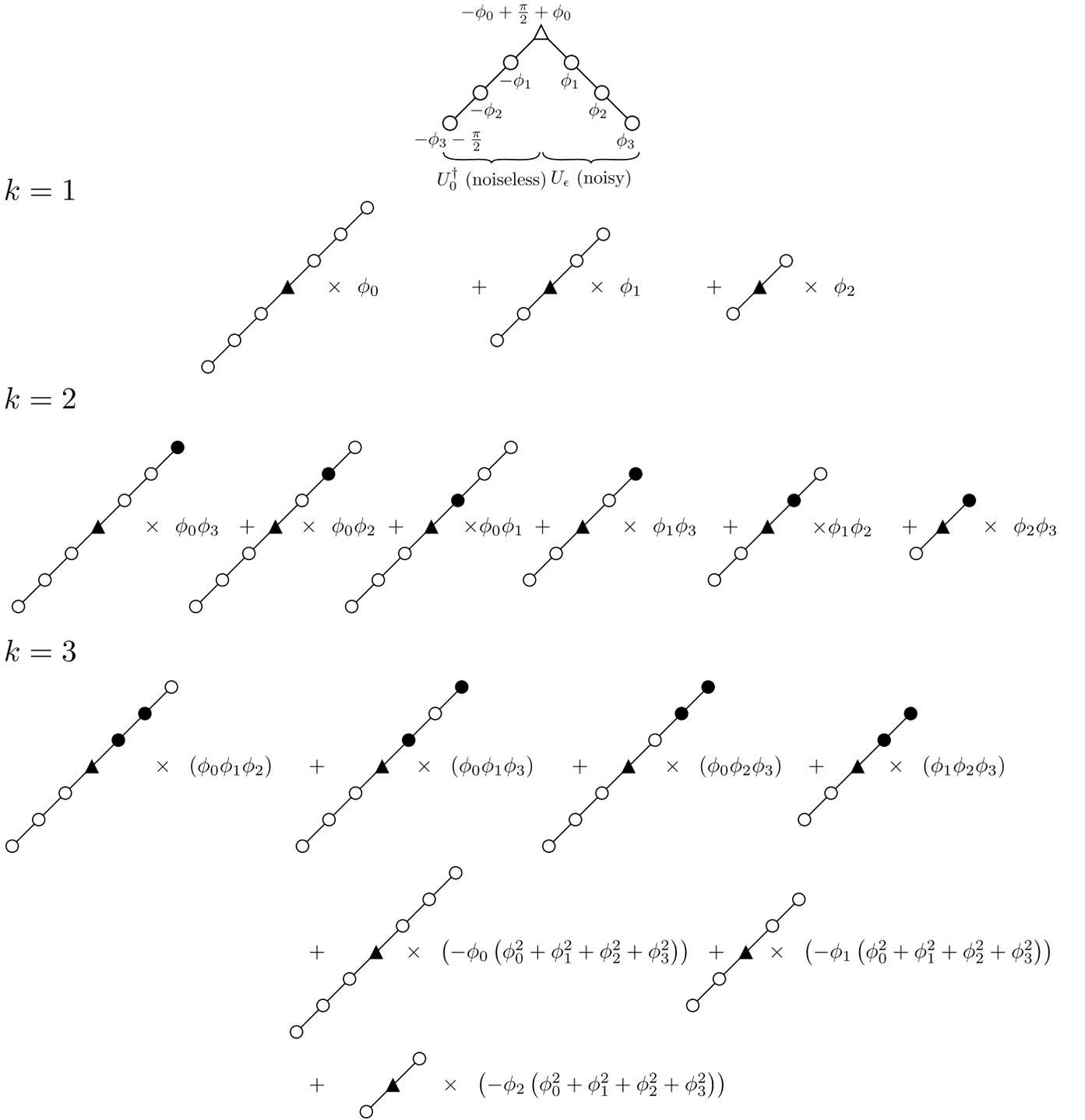

$k = 1$

$k = 2$

$k = 3$

FIG. S5. Diagrammatic representation of error terms for a representative length-3 QSP parameterized by $\phi_i \in \mathbb{R}$ at a) order-1, b) order-2, and c) order-3. To save space, we have omitted phase labels in the expansions. Diagrams are understood to be in the form of Definition S4.6 with open circles denoting $b_i = 0$ and filled circles denoting $b_i = 1$.

where $b_i \in \{0, 1\}$. Furthermore, it is assumed that no $\phi_i$ for $i < d$ is a half-integer multiple of $\pi$; otherwise, we can perform elision to simplify the diagram.

**Remark S4.7** ($\pi$-rotation identity). Let $V$ be a length-$d$ QSP operator parameterized by real phases $(\phi_0, \ldots, \phi_d)$. Then the length-$d$ QSP $V' = \mathrm{QSP}(\theta; (\phi_0, \ldots, \phi_i + \pi, \ldots, \phi_d))$ with single phase over-rotated by $\pi$ is simply related via negation,

$$V = -V'. \tag{S61}$$

This can be clearly seen as a $Z$-rotation by $\pi$ is equivalent to negative identity (i.e $e^{i\pi Z} = -I$), and is shown diagrammatically in Fig. S4 a) for QSP $V$ in the form of Definition S4.6.

**Remark S4.8** ($\pi/2$-rotation identity). Let $V'$ be a length-$2r$ QSP operator parameterized

$$V' = \mathrm{QSP}(\theta; (-\phi_d - \pi/2, \ldots, -\phi_i + \pi/2, \ldots, -\phi_{d-r+1}, \pi, \phi_{d-r+1} + b_{d-r+1}\pi/2, \ldots, \phi_i, \ldots, \phi_d + b_d\pi/2)), \tag{S62}$$

with $\phi_i \in \mathbb{R}$ and $b_i \in \{0, 1\}$ (i.e. nearly in the form Definition S4.6 save for the $\pi/2$-over-rotation of the $-\phi_i$ phase). Using Remark S1.4 and Remark S1.9, we find that

$$V' \sim -V \tag{S63}$$

for $V$ in the form Definition S4.6

$$V = \mathrm{QSP}(\theta; (-\phi_d - \pi/2, \ldots, -\phi_i, \ldots, -\phi_{d-r+1}, \pi, \phi_{d-r+1} + b_{d-r+1}\pi/2, \ldots, \phi_i + \pi/2, \ldots, \phi_d + b_d\pi/2)). \tag{S64}$$

This is shown diagrammatically for an example QSP in Fig. S4 b).

**Remark S4.9** (Standard form, higher-order). To simplify the analysis of error and recovery operators at any order, we generalize Definition S2.3 by writing the contribution at each order as an $XY$-equivalent linear combination of diagrams of the form Definition S4.6. Note that the first-order analysis presented in Remark S4.4 is already in this form. For all higher-orders, this can be accomplished using repeated application of Remark S4.7 and Remark S4.8, keeping track of factors of $-1$.

Diagrams to order $k = 3$ are shown for a generic length-3 QSP in Fig. S5.

### C. First-order recovery revisited

In this subsection, we revisit the construction for first-order recovery of Remark S4.5. In doing so, we improve the scaling from $\Theta(d^2)$ to $\Theta(cd)$, where $c$ is the number of distinct phases (up to factors of $2\pi$).

**Remark S4.10** (First-order component-wise recovery, revisited). We once again consider length-$2r$ recovery operators similar to those in Eq. (S58), but we make use of an additional integral degree of freedom, namely the freedom to over-rotate by factors of $2\pi$ (note that we can also choose to over-rotate to the left of the degree-peak as shown in Fig. S6),

$$V = \mathrm{QSP}(\theta; (-\phi_d - \pi/2, -\phi_{d-1}, \ldots, -\phi_{d-r+1}, \pi/2, \phi_{d-r+1} + 2\pi m_{d-r+1}, \ldots, \phi_d + 2\pi m_d)). \tag{S65}$$

The length-$2r$ recovery operator yields to first-order

$$\begin{aligned}
V = I + \epsilon \Big( &-\frac{n\pi}{2} \times \mathrm{QSP}(\theta; (\pi/2)) \\
&+ 2\pi m_d \times \mathrm{QSP}(\theta; (-\phi_d - \pi/2, \pi, \phi_d))) \\
&+ 2\pi m_{d-1} \times \mathrm{QSP}(\theta; (-\phi_d - \pi/2, -\phi_{d-1}, \pi, \phi_{d-1}, \phi_d))) \\
&\quad \vdots \\
&+ \frac{n\pi}{2} \times \mathrm{QSP}(\theta; (-\phi_d - \pi/2, \ldots, -\phi_{d-r+1}, \pi, \phi_{d-r+1}, \ldots, \phi_d))) \Big) \\
&+ O(\epsilon^2).
\end{aligned} \tag{S66}$$

for some $n \in \mathbb{Z}$ and $m_i \in \mathbb{Z}$ to be specified later. Note that previously in Eq. (S58), we had effectively set $m_i = 0$.

We find the first-order expansion of the recovery operator in Eq. (S66) is now a sum of $r + 1$ diagrams of degree 0 through $2r$ in increments of 2. This is depicted in Fig. S6, and is again nearly identical to the general first-order error term of Fig. S5. The main limitation in the construction of the recovery operator is the real-valued rescaling required to match the generic error term. As in Remark S4.5, this is accomplished using the counter-rotation trick which one can verify generates rescaled $XY$-equivalent diagrams (Fig. S4 c)).

For a generic length-$d$ QSP with all phases distinct, we cannot do better using this method, as we still need to perform counter-rotation $d$ times. But for a length-$d$ QSP operator with $c$ distinct phases, we can group diagrams that are scaled by the same amount together; each group can corrected using a single length-$2r$ diagram by appropriately choosing $m_i$ and $n$. Overall, if there are $c$ distinct phases, there will be $c$ distinct groups, each requiring a separated counter-rotated diagram of length $\Theta(d)$. Note that we count phases as equivalent if they differ only by an integer multiple of $2\pi$ as these can matched within the same group by an appropriate choice of $m_i$ and $n$. Thus the overall complexity using this scheme yields the improved $\Theta(cd)$ for QSP diagrams with high phase degeneracy.

**Remark S4.11** (Example for QSP with single unique phase.). Let $U_\epsilon$ be a length-$d$ QSP with a single unique phase $\phi$ (e.g. Grover search). Using the result of Remark S4.10, we can construct a first-order recovery operator by choosing phases

$$\vec{\eta}_1 = (-\phi - \pi - \delta, -\phi, \ldots, -\phi, \pi/2, \phi + 2\pi m, \ldots, \phi + 2\pi m, \phi + \pi/2 + \delta), \tag{S67}$$

$$\vec{\eta}_2 = (-\phi - 4n\pi - \delta, -\phi, \ldots, -\phi, 4n\pi - \pi/2, \phi + 2\pi m, \ldots, \phi + 2\pi m, \phi + \pi/2 + \delta), \tag{S68}$$

$$\vec{\eta}_3 = (-\phi - \pi + \delta, -\phi, \ldots, -\phi, \pi/2, \phi + 2\pi m, \ldots, \phi + 2\pi m, \phi + \pi/2 - \delta), \tag{S69}$$

$$\vec{\eta}_4 = (-\phi - 4n\pi + \delta, -\phi, \ldots, -\phi, 4n\pi - \pi/2, \phi + 2\pi m, \ldots, \phi + 2\pi m, \phi + \pi/2 - \delta), \tag{S70}$$

$$\tag{S71}$$

with appropriate $m, n \in \mathbb{Z}$ and $\delta$ depending on $\phi$. Note crucially that each $\ldots$ hides only $\Theta(d)$ phases.

Letting $V_\epsilon^{(1)} = \text{QSP}(\theta; \vec{\eta}_1), \ldots, V_\epsilon^{(4)} = \text{QSP}(\theta; \vec{\eta}_1)$, $V_\epsilon^{(1)} V_\epsilon^{(2)} V_\epsilon^{(3)} V_\epsilon^{(4)}$ is a first-order recovery sequence for $U_\epsilon$.

### D. General higher-order recovery

**Remark S4.12** (Constructing higher-order recovery sequences.). Let $R_\epsilon$ and $\bar{R}_\epsilon$ be $k^{\text{th}}$-order unbiased sequences with canonical profiles $\mathcal{R}$ and $\bar{\mathcal{R}}$ respectively with

$$\mathcal{R}_j^{(\sigma,k)} = -\bar{\mathcal{R}}_j^{(\sigma,k)} \quad (\sigma = x, y, z), \tag{S72}$$

for all $j$, and

$$\mathcal{R}_j^{(z,k)} = \bar{\mathcal{R}}_j^{(z,k')} = 0, \tag{S73}$$

for all $j$ for $k' < k$.

Then the operator $S_\epsilon \equiv R_\epsilon \bar{R}_\epsilon$ with canonical profile $\mathcal{S}$ is an order-$(k+1)$ unbiased sequence with

$$\mathcal{S}_j^{(\sigma,k+1)} = \mathcal{R}_j^{(\sigma,k)} + \bar{\mathcal{R}}_j^{(\sigma,k)} \quad (\sigma = x, y, z), \tag{S74}$$

for all $j$.

We can make use of the above observation and Remark S4.4 to generate general higher-order recovery sequences. For instance, to create a second-order recovery sequence, we may combine two first-order sequences. In general, this requires careful choice of $m_i$ and $n$. Example recovery sequences up to order-3 are shown in Fig. S6.

**Lemma S9** (Higher-order component-wise recovery). Let $U_\epsilon^{(k-1)}$ be a noisy QSP unitary $XY$ recovered to order-$(k-1)$,

$$U_\epsilon^{(k-1)} = U_0 e^{i(\chi + O(\epsilon))Z + \epsilon^k((x + O(\epsilon))X + (y + O(\epsilon))Y)}. \tag{S75}$$

If $U_\epsilon^{(k-1)}$ is of length-$d$ and with $c$ unique phases (up to factors of $2\pi$). Then, a recovery sequence $R^{(k)}$ of length $\Theta(2^k c^k d)$ exists to recover up to order-$k$,

$$U_\epsilon^{(k-1)} R^{(k)} = U_0 e^{i(\chi + O(\epsilon))Z + \epsilon^k((x + O(\epsilon))X + (y + O(\epsilon))Y)}. \tag{S76}$$

$k = 1$

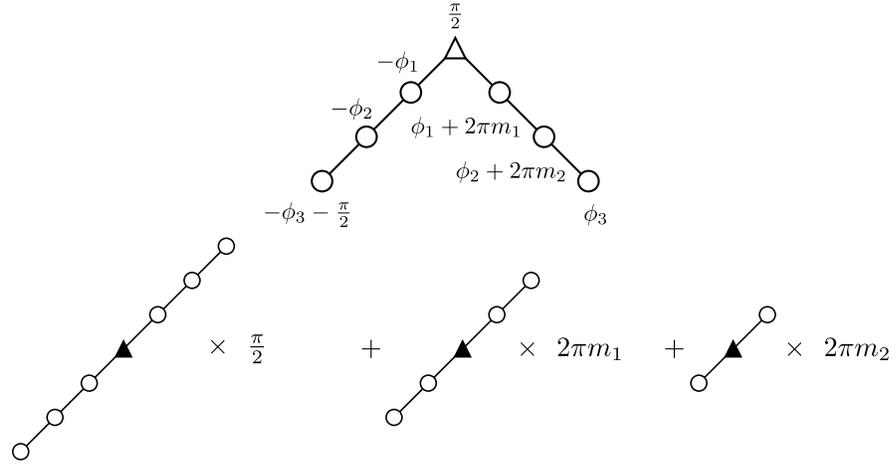

$k = 2$

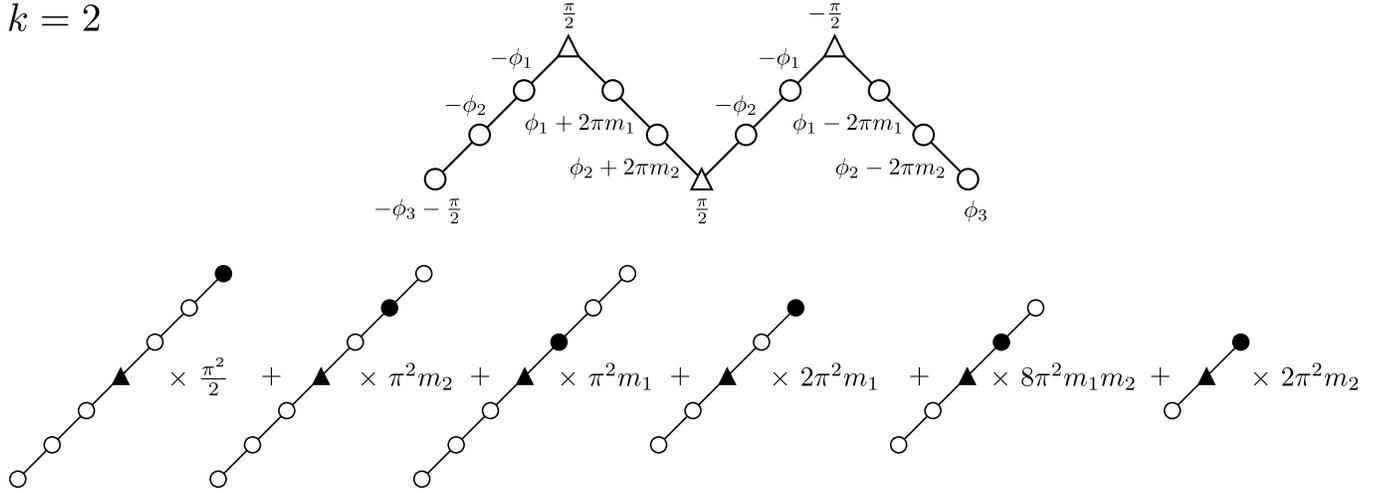

$k = 3$

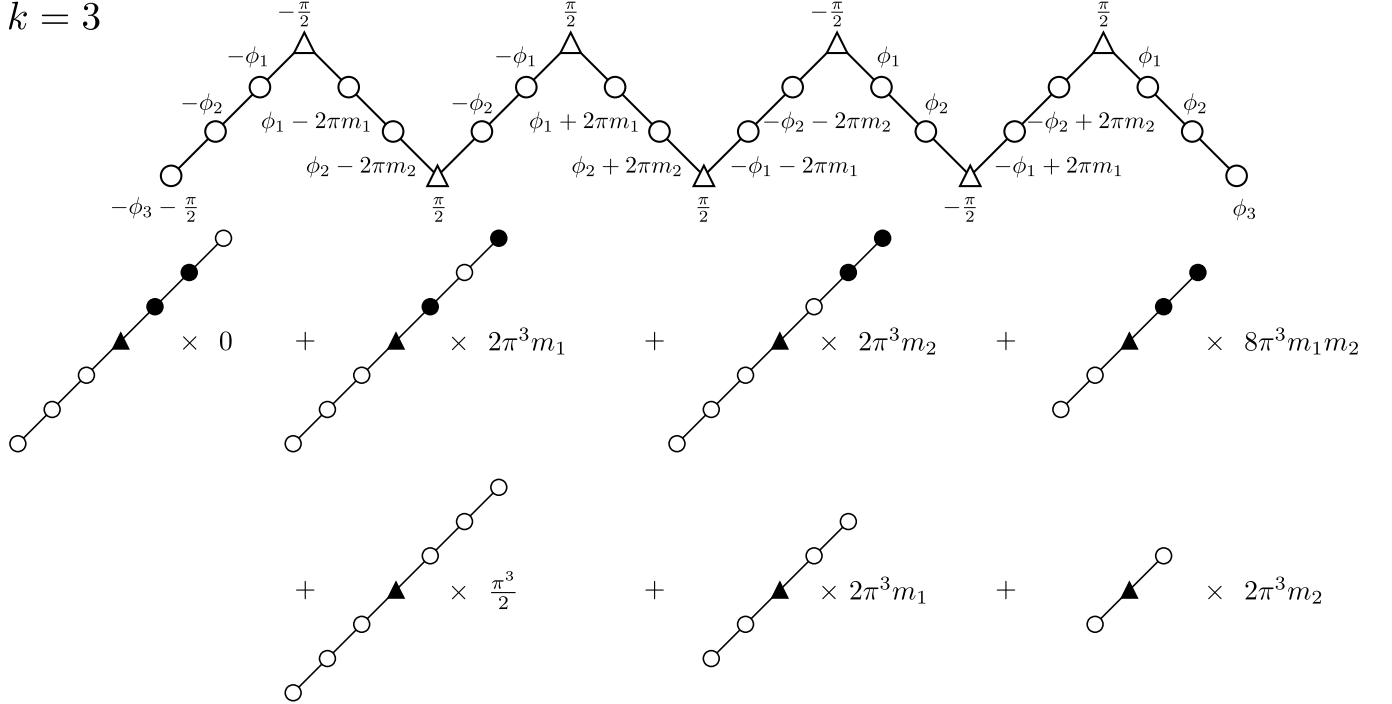

FIG. S6. Diagrammatic analysis of expansions for representative recovery diagrams of a) order-1, b) order-2, and c) order-3. As in Fig. S6, phase labels are omitted and diagrams are understood to be in the form of Definition S4.6 with open circles denoting $b_i = 0$ and filled circles denoting $b_i = 1$.

*Proof.* The result for higher-orders is similar to that in Remark S4.10 for first-order recovery. The units of recovery at order-$k$ can be constructed using Remark S4.12 and are of length $\Theta(2^k d)$ (example recovery units are shown in Fig. S6).

As in the first-order case, we can correct groups of diagrams scaled by the same real coefficient using each recovery unit. In general, each recovery unit may need to be repeated a constant number of times with different values of $m_i$ and $n$ in order to attain the desired integral coefficients.

The main factor bounding recovery length is the number of distinct groups with different real coefficients. The real coefficients at order-$k$ consist of the $k$-tuples of the $c$ distinct phases, of which there are $\binom{c}{k} = \Theta(c^k)$. Note that again we consider phases equivalent if they differ only by an integer multiple of $2\pi$ as these require only $o(c^k)$ additional recovery units.

Thus the recovery to order-$k$ can be accomplished using a recovery sequence of length $\Theta(2^k c^k d)$. □

We are now ready to prove Theorem 3, restated below, in full generality.

**Theorem 3** (Upper bound on recovery length). *Given any noisy QSP operator $U_\epsilon(\theta)$ of length $d$ with $c$ distinct phases (up to factors of $2\pi$) and an integer $k \geq 1$, there exists a recovery sequence $R_\epsilon(\theta)$ satisfying*

$$|\langle 0|U_\epsilon R_\epsilon|0\rangle|^2 = |\langle 0|U_0|0\rangle|^2 + O(\epsilon^{k+1})$$

*for all $\theta$. Furthermore, there exists a QSP operator satisfying the above with length at most $O(2^k c^{k^2} d)$.*

*Proof.* Let $U_\epsilon$ be a length-$d$ QSP sequence with $c$ unique phases with error operator $E_\epsilon$. We perform recovery order-by-order so that we can make use of the additive property of diagrams to leading-order (Remark S2.5). At each order, from Remark S4.1, it suffices to append a sequence of relatively negative $XY$-equivalent diagrams. Recovery to first-order has been shown in Remark S4.10 with a sequence $R_\epsilon^{(1)}$ of length $\Theta(cd)$.

After appending the first-order recovery sequence, we have $U_\epsilon R_\epsilon^{(1)}$, a length $\Theta(cd)$ QSP sequence. The key to showing the desired scaling is to notice this sequence has large phase redundancy: namely, the length $\Theta(cd)$ QSP sequence is parameterized by the same phases, except for an additional $\Theta(c)$ new distinct phases used in the counter-rotation. Thus, by Lemma S9, recovery to second-order can be accomplished using a recovery sequence of length $\Theta(2^2 \times c^2 \times cd) = \Theta(2^2 c^3 d)$. Recovery to order-$k$ can be accomplished using a recovery sequence a factor $\Theta(c^k)$ longer than the previous order. Overall this construction requires a length $O(2^k c^{k^2} d)$ sequence, thus proving the Theorem. □

We have Theorem 2, restated below, as a corollary.

**Theorem 2** (Recoverability). *Given any noisy QSP operator $U_\epsilon(\theta)$ of length $d$ and an integer $k \geq 1$, there exists a recovery sequence $R_\epsilon(\theta)$ satisfying*

$$|\langle 0|U_\epsilon R_\epsilon|0\rangle|^2 = |\langle 0|U_0|0\rangle|^2 + O(\epsilon^{k+1}) \tag{S77}$$

*for all $\theta$.*

In Section S5, we provide an alternate proof to Theorem 2.

## S5. Alternate Proof of Theorem 2: Degree-wise Recovery

In this Appendix, we present an alternate construction for recovery. Though this construction is exponentially less efficient, generating a length $\Theta(2^k d^{2^k})$ sequence, we find that it is sufficiently different to be worth discussion. Furthermore, we find that despite being asymptotically worse, it can produce shorter recovery sequences in practice due to the large constants hidden in Lemma S9. Whereas the construction presented in performs recovery component-wise, here we perform recovery degree-wise.

### A. First-order

We now describe our construction of the recovery sequence that corrects the first-order error in a QSP sequence. Given a faulty QSP sequence $U_\epsilon$, let $E_\epsilon$ be its error operator and $\mathcal{P}$ its error profile, it suffices by Remark S4.1 to construct a recovery sequence $R_\epsilon$ with an error profile $\mathcal{R}$ satisfying

$$\mathcal{R}_j^{(\sigma,1)} = -\mathcal{P}_j^{(\sigma,1)} \tag{S78}$$

for $\sigma = x, y$ and for all $j$.

The construction of $R_\epsilon$ is recursive. In the first iteration, we construct $R_\epsilon$ that satisfies Eq. (S78) only at $j_{\max}$, the largest $j$ such that $\mathcal{P}_j^{(\sigma,1)} \neq 0$ ($\sigma = x, y$). At the end of this iteration, the appended QSP sequence has a modified error profile $\mathcal{P}'$ such that $\mathcal{P}_j'^{(\sigma,0)} = 0$ for all $j \geq j_{\max}$, resulting in a lower $j_{\max}$ for the next iteration. Repeating this procedure until $\mathcal{P}_j^{(\sigma,0)} = 0$ for all $j$, we arrive at the desired recovery sequence.

The building block for our recovery sequence is the conjugations in Eq. (S56). The following Lemma gives the error profile for QSP sequence that results from the conjugations.

**Lemma S10** (Top-degree recovery term, first-order). *Let $R$ be the length-$(2d)$ QSP sequence resulting from $d$ conjugations in Eq. (S56) with $m_1 = \cdots = m_d = 0$ and $n_1 = \cdots = n_d = n$. Let $\mathcal{R}$ be the error profile of $R_\epsilon$. We have*

$$\begin{pmatrix} \mathcal{R}_{d-1}^{(x,1)} \\ \mathcal{R}_{d-1}^{(y,1)} \\ \mathcal{R}_{d-1}^{(z,1)} \end{pmatrix} = \pi 2^{2d-3}(2n+1) \prod_{j=1}^{d-1} \cos^2(\eta_j) \begin{pmatrix} \sin(2\eta_d) \\ -\cos(2\eta_d) \\ 2 \end{pmatrix}. \tag{S79}$$

*Proof.* We prove Lemma S10 by induction. For $d = 1$, the error profile of the corresponding length-2 QSP satisfies Eq. (S79):

$$\begin{pmatrix} \mathcal{R}_0^{(x,1)} \\ \mathcal{R}_0^{(y,1)} \\ \mathcal{R}_0^{(z,1)} \end{pmatrix} = \pi 2^{-1}(2n+1) \begin{pmatrix} \sin(2\eta_1) \\ -\cos(2\eta_1) \\ 2 \end{pmatrix}. \tag{S80}$$

Suppose Lemma S10 holds for all length-$2d$ sequences. We shall prove that it also holds for all length-$(2d + 2)$ sequences.

Let $R'$ be a length-$(2d + 2)$ sequence satisfying the assumptions of Lemma S10 and $R$ be a length-$2d$ sequence satisfying

$$R' = \mathcal{C}_{0,n,\eta_{d+1}} R. \tag{S81}$$

Let $\mathcal{R}'$ and $\mathcal{R}$ be the error profiles of $R'_\epsilon$ and $R_\epsilon$, respectively. Using Eq. (S22), we have

$$\mathcal{R}_d'^{(x,0)} = \sin(2\eta_{d+1})(\mathcal{R}_{d-1}^{(z,0)} - 2\mathcal{R}_{d-1}^{(y,0)}), \tag{S82}$$

$$\mathcal{R}_d'^{(y,0)} = -\cos(2\eta_{d+1})(\mathcal{R}_{d-1}^{(z,0)} - 2\mathcal{R}_{d-1}^{(y,0)}), \tag{S83}$$

$$\mathcal{R}_d'^{(z,0)} = 2(\mathcal{R}_{d-1}^{(z,0)} - 2\mathcal{R}_{d-1}^{(y,0)}). \tag{S84}$$

Applying Lemma S10 on $\mathcal{R}$, we have

$$\mathcal{R}_{d-1}^{(z,0)} - 2\mathcal{R}_{d-1}^{(y,0)} = \pi 2^{2d-3}(2n+1)\prod_{j=1}^{d-1}\cos^2(\eta_j)(2+2\cos(2\eta_d)) \tag{S85}$$

$$= \pi 2^{2d-1}(2n+1)\prod_{j=1}^{d}\cos^2(\eta_j). \tag{S86}$$

Therefore,

$$\begin{pmatrix} \mathcal{R}_d'^{(x,1)} \\ \mathcal{R}_d'^{(y,1)} \\ \mathcal{R}_d'^{(z,1)} \end{pmatrix} = \pi 2^{2d-1}(2n+1)\prod_{j=1}^{d}\cos^2(\eta_j)\begin{pmatrix} \sin(2\eta_{d+1}) \\ -\cos(2\eta_{d+1}) \\ 2 \end{pmatrix} \tag{S87}$$

and Lemma S10 holds for $R'$. By induction, Lemma S10 holds for length-$2d$ QSP sequences for all $d \geq 1$. □

Also, recall from Remark S1.9 that $R_j^{(x,1)} = R_j^{(y,1)} = R_j^{(z,1)} = 0$ for all $j \geq d$.

**Lemma S11** (First-order degree-wise recovery). *Let $U$ be a length-$d$ QSP sequence, $E_\epsilon$ be the error operator for $U$, and $\mathcal{P}$ its error profile. Let $j_{\max}$ be the largest $j$ such that either $\mathcal{P}_j^{(x,1)} \neq 0$ or $\mathcal{P}_j^{(y,1)} \neq 1$. There exists an unbiased recovery sequence $R$ such that the error profile $\mathcal{P}'$ of $U_\epsilon R_\epsilon$ satisfies*

$$\mathcal{P}_j'^{(x,1)} = \mathcal{P}_j'^{(y,1)} = 0 \tag{S88}$$

*for all $j \geq j_{\max}$. In addition, the length of the recovery sequence is at most $2(j_{\max}+1)$ if $j_{\max} \geq 1$ and at most 4 if $j_{\max} = 0$.*

*Proof.* First, we consider $j_{\max} \geq 1$. Let $n$ be the smallest integer such that

$$\sqrt{(\mathcal{P}_{j_{\max}}^{(x,1)})^2 + (\mathcal{P}_{j_{\max}}^{(y,1)})^2} \leq \pi 2^{2j_{\max}-1}\left(n+\frac{1}{2}\right). \tag{S89}$$

Let $R$ be the length-$(2j_{\max}+2)$ QSP sequence in Eq. (S56) with $m_1 = \cdots = m_{j_{\max}+1} = 0$, $n_1 = \cdots = n_{j_{\max}+1} = n$ and $\mathcal{R}$ be the error profile of $R_\epsilon$. By Lemma S10, we have

$$\begin{pmatrix} \mathcal{R}_{j_{\max}}^{(x,1)} \\ \mathcal{R}_{j_{\max}}^{(y,1)} \end{pmatrix} = \pi 2^{2j_{\max}-1}\left(n+\frac{1}{2}\right)\prod_{j=1}^{j_{\max}}\cos^2(\eta_j)\begin{pmatrix} \sin(2\eta_{j_{\max}+1}) \\ -\cos(2\eta_{j_{\max}+1}) \end{pmatrix}. \tag{S90}$$

Next, we choose $\eta_1 = \cdots = \eta_{j_{\max}-1} = 0$,

$$\eta_{j_{\max}} = \cos^{-1}\left(\frac{\sqrt{(\mathcal{P}_{j_{\max}}^{(x,1)})^2 + (\mathcal{P}_{j_{\max}}^{(y,1)})^2}}{\pi 2^{2j_{\max}-1}\left(n+\frac{1}{2}\right)}\right)^{1/2} \tag{S91}$$

$$\eta_{j_{\max}+1} = -\frac{1}{2}\tan^{-1}\left(\frac{\mathcal{P}_{j_{\max}}^{(x,1)}}{\mathcal{P}_{j_{\max}}^{(y,1)}}\right) \leq 0. \tag{S92}$$

Substituting these phase angles into Eq. (S90), we obtain

$$\begin{pmatrix} \mathcal{R}_{j_{\max}}^{(x,1)} \\ \mathcal{R}_{j_{\max}}^{(y,1)} \end{pmatrix} = -\begin{pmatrix} \mathcal{P}_{j_{\max}}^{(x,1)} \\ \mathcal{P}_{j_{\max}}^{(y,1)} \end{pmatrix}. \tag{S93}$$

Thus by Remark S2.5 we have the Lemma for $j_{\max} \geq 1$.

Finally, we consider the case $j_{\max} = 0$. Recall that for $j_{\max} \geq 1$, we have a continuous control over the magnitude of the error profile provided by $\eta_{j_{\max}}$. For $j_{\max} = 0$, we use the counter-rotation trick of Remark S4.3. Accordingly,

we choose $\eta = -\tan^{-1}\left(\mathcal{P}_0^{(x,1)}/\mathcal{P}_0^{(y,1)}\right) \leq 0$ and

$$\delta\eta = \frac{1}{2}\cos^{-1}\left(\frac{\sqrt{(\mathcal{P}_0^{(x,1)})^2 + (\mathcal{P}_0^{(y,1)})^2}}{\pi\left(n + \frac{1}{2}\right)}\right) \tag{S94}$$

to arrive at Eq. (S93) for $j_{\max} = 0$. This concludes the proof of the Lemma. $\qquad\square$

Repeatedly applying Lemma S11, we incrementally lower $j_{\max}$. When $\mathcal{P}_j^{(x,0)} = \mathcal{P}_j^{(y,0)} = 0$ for all $j \geq 0$, we arrive at Theorem 2 for $k = 1$. Since the length of the recovery sequence in each iteration of Lemma S10 is $2(j_{\max} + 1)$ and $j_{\max}$ is initially at most $d - 1$, the total length of the recovery sequence is at most

$$4 + \sum_{j_{\max}=1}^{d-1} 2(j_{\max} + 1) = d^2 + d + 2. \tag{S95}$$

We arrive at the following corollary:

**Corollary S12** (First-order recovery length). *For a length-$d$ QSP $U_\epsilon$, there exists recovery sequence $R_\epsilon$ of length $d^2 + d + 2$ satisfying Theorem 2 for $k = 1$, that is*

$$|\langle 0|U_\epsilon R_\epsilon|0\rangle|^2 = |\langle 0|U_0|0\rangle|^2 + O(\epsilon^2). \tag{S96}$$

### B. Higher-order

We now generalize to higher-orders. First, we provide an explicit recursive construction of higher order unbiased sequences.

**Lemma S13** (Top-degree recovery term, higher-order). *For all $k \geq 1$, there exists a $k^{\text{th}}$-order unbiased QSP sequence of length-$(2^k d)$ $R_\epsilon$, parameterized by $\eta_1, \ldots, \eta_d \in [-\pi, \pi)$ and $n \in \mathbb{Z}$ with an error profile $\mathcal{R}$ satisfying*

$$\begin{pmatrix} \mathcal{R}_{d-1}^{(x,k)} \\ \mathcal{R}_{d-1}^{(y,k)} \end{pmatrix} = \pi^k 2^{2d-3}(2n+1)^k \prod_{j=1}^{d-1}\cos^2(\eta_j) \begin{pmatrix} 0 & 1 \\ -1 & 0 \end{pmatrix}^{k-1} \begin{pmatrix} \sin(2\eta_d) \\ -\cos(2\eta_d) \end{pmatrix} \tag{S97}$$

*and $\mathcal{R}_j^{(x,k-1)} = \mathcal{R}_j^{(y,k-1)} = 0$ for all $j \geq d$.*

*Proof.* We will provide a recursive construction for a length-$(2^k d)$ QSP satisfying the Lemma.

Let $R_\epsilon$ be the length-$2d$ recovery sequence parameterized by $n \in \mathbb{Z}$ and $\eta_1, \ldots, \eta_d$ as in Lemma S10 and $\mathcal{R}$ its error profile. From the Lemma, we have

$$\begin{pmatrix} \mathcal{R}_{d-1}^{(x,1)} \\ \mathcal{R}_{d-1}^{(y,1)} \end{pmatrix} = \pi 2^{2d-3}(2n+1)\prod_{j=1}^{d-1}\cos^2(\eta_j) \begin{pmatrix} 0 & 1 \\ -1 & 0 \end{pmatrix}^0 \begin{pmatrix} \sin(2\eta_d) \\ -\cos(2\eta_d) \end{pmatrix}, \tag{S98}$$

and $\mathcal{R}_j^{(x,0)} = \mathcal{R}_j^{(y,0)} = 0$ for all $j \geq d$. So $R$ satisfies the Lemma for $k = 1$.

Suppose the Lemma holds for $k \geq 1$ and let $R_\epsilon$ be the $k^{\text{th}}$-order unbiased QSP sequence satisfying the Lemma. We define $\bar{R}_\epsilon$ to be the QSP sequence identical to $R_\epsilon$ except that $\eta_d \mapsto \eta_d + \frac{\pi}{2}$ and $n \mapsto -(n+1)$. Let $\mathcal{R}$ and $\bar{\mathcal{R}}$ be their respective error profiles. Using Remark S4.12, one can show that

$$\mathcal{R}_j^{(\sigma,k)} = \bar{\mathcal{R}}_j^{(\sigma,k)} \quad (\sigma = x, y), \tag{S99}$$

$$\mathcal{R}_j^{(z,k)} = -\bar{\mathcal{R}}_j^{(z,k)}, \tag{S100}$$

for all $j$ and $k \geq 1$.

Thus we can construct a sequence $S$ using Remark S4.12 that is unbiased to order $k + 1$ as follow:

$$S_\epsilon \equiv e^{-i\pi(n+\frac{1}{2})(1+\epsilon)Z} R_\epsilon e^{i\pi(n+\frac{1}{2})(1+\epsilon)Z} \bar{R}_\epsilon, \tag{S101}$$

which is a length-$(2^{k+1}d)$ QSP unitary. By the result of Remark S4.12, the error profile $\mathcal{S}$ of $S_\epsilon$ satisfies

$$\begin{pmatrix} \mathcal{S}_{d-1}^{(x,k)} \\ \mathcal{S}_{d-1}^{(y,k)} \end{pmatrix} = \pi(2n+1)\begin{pmatrix} 0 & 1 \\ -1 & 0 \end{pmatrix}\begin{pmatrix} \mathcal{R}_{d-1}^{(y,k-1)} \\ \mathcal{R}_{d-1}^{(x,k-1)} \end{pmatrix} = \pi^k 2^{2d-3}(2n+1)^k \prod_{j=1}^{d-1} \cos^2(\eta_j) \begin{pmatrix} 0 & 1 \\ -1 & 0 \end{pmatrix}^{k-1} \begin{pmatrix} \sin(2\eta_d) \\ -\cos(2\eta_d) \end{pmatrix}. \quad (S102)$$

Therefore the Lemma holds for $k+1$ and, by induction, it holds for all $k$. $\qquad\square$

**Lemma S14** (Higher-order degree-wise recovery). *Let $U$ be a length-$d$ QSP sequence, $E_\epsilon$ its error operator, and $\mathcal{P}$ its error profile. Suppose $E_\epsilon$ is unbiased to order-$k$, that is $\mathcal{P}_j^{(x,k')} = \mathcal{P}_j^{(y,k')} = 0$ for all $j \geq 0$ and $k' < k$. Let $j_{\max}$ be the largest $j$ such that either $\mathcal{P}_j^{(x,k)} \neq 0$ or $\mathcal{P}_j^{(y,k)} \neq 0$. There exists an unbiased recovery sequence $R$ such that the error profile $\mathcal{P}'$ of $U_\epsilon R_\epsilon$ satisfies*

$$\mathcal{P}_j'^{(x,k)} = \mathcal{P}_j'^{(y,k)} = 0 \quad (S103)$$

*for all $j \geq j_{\max}$ and $k \geq 0$. In addition, the length of the recovery sequence is at most $2^{k+1}(j_{\max}+1)$ if $j_{\max} \geq 1$ and at most $2^{k+2}$ if $j_{\max} = 0$.*

*Proof.* The proof of the Lemma is nearly identical to that of Lemma S11.

First, consider $k \geq 1$ and $j_{\max} \geq 1$. Let $n$ be the smallest integer such that

$$\sqrt{(\mathcal{P}_{j_{\max}}^{(x,k)})^2 + (\mathcal{P}_{j_{\max}}^{(y,k)})^2} \leq \pi^k 2^{2j_{\max}-1}(2n+1)^k. \quad (S104)$$

Let $R$ be the $k^{\text{th}}$-order unbiased length-$(2^{k+1}(j_{\max}+1))$ QSP sequence parameterized by $\eta_1, \ldots, \eta_{j_{\max}+1} \in [-\pi, \pi)$ and $n \in \mathbb{Z}$, that satisfies Lemma S13 and $\mathcal{R}$ be its error profile. From Lemma S13, we have

$$\begin{pmatrix} \mathcal{R}_{j_{\max}}^{(x,k)} \\ \mathcal{R}_{j_{\max}}^{(y,k)} \end{pmatrix} = \pi^k 2^{2j_{\max}-1}(2n+1)^k \prod_{j=1}^{j_{\max}} \cos^2(\eta_j)\begin{pmatrix} 0 & 1 \\ -1 & 0 \end{pmatrix}^{k-1} \begin{pmatrix} \sin(2\eta_{j_{\max}+1}) \\ -\cos(2\eta_{j_{\max}+1}) \end{pmatrix}, \quad (S105)$$

Next, we choose $\eta_1 = \cdots = \eta_{j_{\max}-1} = 0$, $n$ from Eq. (S104), and

$$\eta_{j_{\max}} = \cos^{-1}\left(\frac{\sqrt{(\mathcal{P}_{j_{\max}}^{(x,k)})^2 + (\mathcal{P}_{j_{\max}}^{(y,k)})^2}}{\pi^k 2^{2j_{\max}-1}(2n+1)^k}\right)^{1/2} \quad (S106)$$

$$\eta_{j_{\max}+1} = -\frac{1}{2}\tan^{-1}\left(\frac{\mathcal{P}_{j_{\max}}^{(x,k)}}{\mathcal{P}_{j_{\max}}^{(y,k)}}\right) - \frac{3\pi k}{4}. \quad (S107)$$

Substituting these phase angles into Eq. (S105), we obtain

$$\begin{pmatrix} \mathcal{R}_{j_{\max}}^{(x,k)} \\ \mathcal{R}_{j_{\max}}^{(y,k)} \end{pmatrix} = -\begin{pmatrix} \mathcal{P}_{j_{\max}}^{(x,k)} \\ \mathcal{P}_{j_{\max}}^{(y,k)} \end{pmatrix}. \quad (S108)$$

From Remark S2.5, we have Lemma S14 for $j_{\max} \geq 1$.

Finally, for the case $j_{\max} = 0$ we again use the counter-rotation trick of Remark S4.3 setting $\eta_1 = \eta \pm \delta\eta$. We choose $\eta = -\tan^{-1}\left(\mathcal{P}_0^{(x,0)}/\mathcal{P}_0^{(y,0)}\right) - \frac{3\pi k}{4}$ and

$$\delta\eta = \frac{1}{2}\cos^{-1}\left(\frac{2\sqrt{(\mathcal{P}_0^{(x,k)})^2 + (\mathcal{P}_0^{(y,k)})^2}}{\pi^k(2n+1)^k}\right) \quad (S109)$$

to arrive at Eq. (S108) for $j_{\max} = 0$. This concludes the proof of Lemma S14. $\qquad\square$

This provides an alternate proof for Theorem 2 which is restated below for convenience.

**Theorem 2** (Recoverability). *Given any noisy QSP operator $U_\epsilon(\theta)$ of length $d$ and an integer $k \geq 1$, there exists a recovery sequence $R_\epsilon(\theta)$ satisfying*

$$|\langle 0|U_\epsilon R_\epsilon|0\rangle|^2 = |\langle 0|U_0|0\rangle|^2 + O(\epsilon^{k+1}) \tag{S110}$$

*for all $\theta$.*

*Proof.* To prove Theorem 2 in full generality for $k \geq 1$, we repeatedly apply Lemma S14.

Let $R_\epsilon^{(k')}$ be the recovery operator accumulated from such repeated applications for order-$k'$. We start by constructing $R_\epsilon^{(1)}$ such that $U_0^\dagger U_\epsilon R_\epsilon^{(1)}$ is unbiased to order-2. We then increment $k'$, repeating the process up to $k' = k+1$ to obtain a $(k+1)^{\text{th}}$-order unbiased operator $U_0^\dagger U_\epsilon R_\epsilon^{(1)} \ldots R_\epsilon^{(k)}$.

We can therefore write

$$U_\epsilon R_\epsilon^{(1)} \ldots R_\epsilon^{(k)} \equiv U_0 e^{i\epsilon(z+O(\epsilon))Z + i\epsilon^{k+1}[(x+O(\epsilon))X + (y+O(\epsilon))Y]}, \tag{S111}$$

as required by Remark S4.1 thus providing an alternate proof of Theorem 2. □

**Corollary S15** (Higher-order recovery length). *For a length-$d$ QSP $U_\epsilon$, there exists a $XY$-recovered sequence $U_\epsilon'$ of length $\Theta(2^k d^{2^k})$ satisfying Remark S4.1 for all $k \geq 1$.*

*Proof.* The proof is by induction on $k$.

For $k = 1$, Corollary S12 gives the exact length of $d^2 + d + 2 = \Theta(2^1 d^{2^1})$ which agrees with Theorem 3.

We now show the inductive step assuming the Corollary holds at order $k$. Given a noisy QSP corrected to order $k$ of length-$d_k$ with $d_k = \Theta(2^k d^{2^k})$, we can perform correction to the $(k+1)^{\text{th}}$-order using at most $d_k$ applications of Lemma S13 for each $0 \leq j_{\max} \leq d_k - 1$. From Lemma S14, each application adds length $2^k(j_{\max} + 1)$ for $j_{\max} > 0$ and $2^{k+1}$ for $j_{\max} = 0$. The overall length of the resulting sequence therefore has length $d_k$ satisfying

$$d_{k+1} = d_k + 2^{k+2} + \sum_{j_{\max}=1}^{d_k} 2^{k+1}(j_{\max} + 1) = \Theta(2^{k+1} d^{2^{k+1}}). \tag{S112}$$

Thus Theorem 3 holds for all $k \geq 1$. □

The main reason for the exponentially worse performance as compared with the construction in Section S4 is the fact that we do not make use of the phase redundancy in the recovered operators, and thus recovery at each order is performed de novo.

## S6. Proof of Theorem 4: Lower Bound

In this Appendix, we show that the length of our recovery sequence for first-order recovery is asymptotically optimal.

While the assumption on the first-order $Z$ component in Theorem 4 is required for technical reasons, the condition can also be seen as a desire to limit the complexity of the recovery sequence. It is important to point out that the condition for $XY$ recovery (Remark S4.1) does not itself place any limits on $f(a)$; and, in fact, neither the component-wise nor the degree-wise recovery constructions we've presented satisfy this requirement, requiring $f(a) = \Omega(d)$.

First, we define the inverse of the conjugation operation of Remark S1.9.

**Remark S6.1** (Recurrence under anti-conjugation, first-order)**.** We can construct the inverse to the conjugation operation using Corollary S6.

$$\mathcal{C}_{n,\eta}^{-1}\text{QSP}(\theta;\vec{\phi}) \equiv e^{i\pi(n+\frac{1}{2})Z}We^{i\eta Z}\text{QSP}(\theta;\vec{\phi})e^{-i(\eta+\frac{\pi}{2})Z}We^{i\frac{\pi}{2}Z},$$ (S113)

Let $U_\epsilon$ be a degree-$d$ operator with canonical expansion $\mathcal{P}$, the canonical expansion of $\mathcal{C}_{n,\eta}^{-1}U_\epsilon$ is

$$\begin{pmatrix} \vec{\mathcal{P}}'^{(1)}_d \\ \vec{\mathcal{P}}'^{(1)}_{d-1} \\ \vdots \\ \vec{\mathcal{P}}'^{(1)}_0 \\ \vec{\mathcal{P}}'^{(1)}_{-1} \end{pmatrix} = \begin{pmatrix} A(\eta) & & & & \\ B(\eta) & A(\eta) & & & \\ & B(\eta) & A(\eta) & & \\ & & & \ddots & \\ & & & B(\eta) & A(\eta) \\ & & & & B(\eta) \end{pmatrix} \begin{pmatrix} \vec{\mathcal{P}}^{(1)}_{d-1} \\ \vdots \\ \vec{\mathcal{P}}^{(1)}_0 \\ \vec{\mathcal{P}}^{(1)}_{-1} \end{pmatrix}$$ (S114)

where the matrix on the right-hand side is a block bidiagonal matrix of size $4(d+2) \times 4(d+1)$ with blocks

$$A(\eta) \equiv \begin{pmatrix} 0 & -4\sin 2\eta & -4\cos 2\eta & -2 \\ 0 & -2\sin 2\eta & 2\cos 2\eta & 1 \\ 0 & 0 & 0 & 0 \\ 0 & 0 & 0 & 0 \end{pmatrix}, \quad B(\eta) \equiv \begin{pmatrix} 0 & -4\sin 2\eta & 4\cos 2\eta & 1 \\ 0 & \sin 2\eta & -\cos 2\eta & 0 \\ 0 & -\cos 2\eta & -\sin 2\eta & 0 \\ 1 & 0 & 0 & 0 \end{pmatrix}.$$ (S115)

**Lemma S16** (Uniqueness of QSP parameterization)**.** Let $U = \text{QSP}(\theta;\vec{\phi})$ be a length-$d$ QSP and let $V = \text{QSP}(\theta;\vec{\psi})$ be a length-$d'$ QSP. Further assume such that no phase $\phi_i$, $\psi_j$ is a half-integer multiple of $\pi$. Then $U = e^{i\chi}V$ for some global phase $\chi \in [0, 2\pi)$ if and only if $d = d'$ and for all $0 \le i \le d$, $\psi_i - \phi_i = \pi n_i$ for some $n_i \in \mathbb{Z}$. Furthermore, either $\chi = 0$ or $\chi = \pi$.

*Proof.* The $\impliedby$ direction is a straightforward application of Remark S4.7 and so we focus on $\implies$.

Since by assumption, neither $\vec{\phi}$ nor $\vec{\psi}$ contain a half-integer multiple of $\pi$, Lemma S4 implies that neither contain any degree peaks and therefore $\deg(P_U) = d$ and $\deg(P_V) = d'$. As a result, the QSP unitaries must be of the same length $U = V \implies P_U = P_V \implies \deg(P_U) = \deg(P_V) \implies d = d'$. We can therefore limit our consideration to the case of $d = d'$.

Now we show that the phases must be equivalent up to an integer multiple of $\pi$ inductively. First consider the case where $d = 0$. In this case, $U = e^{i\chi}V \implies e^{i\phi_0 Z} = e^{i\chi}e^{i\psi_0 Z} \implies \phi_0 = \psi_0 + \pi n_0$ for some $n_0 \in \mathbb{Z}$; furthermore, $\chi = 0$ for $n_0$ even and $\chi = \pi$ for $n_0$ odd. Thus the Lemma is satisfied for $d = 0$.

Assuming the Lemma for QSP unitaries of length-$d$, we show that it holds for QSP unitaries of length-$(d+1)$. Consider QSPs $U = \text{QSP}(\theta;\vec{\phi})$ and $V = \text{QSP}(\theta;\vec{\psi})$ each of length-$(d+1)$ satisfying the conditions of the Lemma. Given that $U = V$, we reduce $U$ to a length-$d$ QSP by right-multiplying both sides by a QSP inverse (Corollary S6) of its final signal processing step $(We^{i\phi_{d+1}Z})^{-1} = e^{-i(\phi_{d+1}-\frac{\pi}{2})Z}We^{-i\frac{\pi}{2}Z}$. The result is

$$U = V,$$ (S116)

$$\implies \text{QSP}(\theta;\vec{\phi}_{0:d+1}) = \text{QSP}(\theta;\vec{\psi}_{0:d+1}),$$ (S117)

$$\implies \text{QSP}(\theta;\vec{\phi}_{0:d+1})e^{-i(\phi_{d+1}-\frac{\pi}{2})Z}We^{-i\frac{\pi}{2}Z} = \text{QSP}(\theta;\vec{\psi}_{0:d+1})e^{-i(\phi_{d+1}-\frac{\pi}{2})Z}We^{-i\frac{\pi}{2}Z},$$ (S118)

$$\implies \text{QSP}(\theta;\vec{\phi}_{0:d}) = \text{QSP}(\theta;\{\psi_0, \dots, \psi_d, \psi_{d+1} - \phi_{d+1} + \pi/2, -\pi/2\}).$$ (S119)

On the left-hand side of the final equation is a QSP unitary of degree-$d$; by the inductive hypothesis, it must be the case that the QSP unitary on the right-hand side, which is of length-$(d+2)$, is also of degree-$d$. This is only possible if we can perform elision at the next-to-last position. By Lemma S4 this requires $\psi_{d+1} - \phi_{d+1} + \frac{\pi}{2} = \pi(n_{d+1} + \frac{1}{2})$

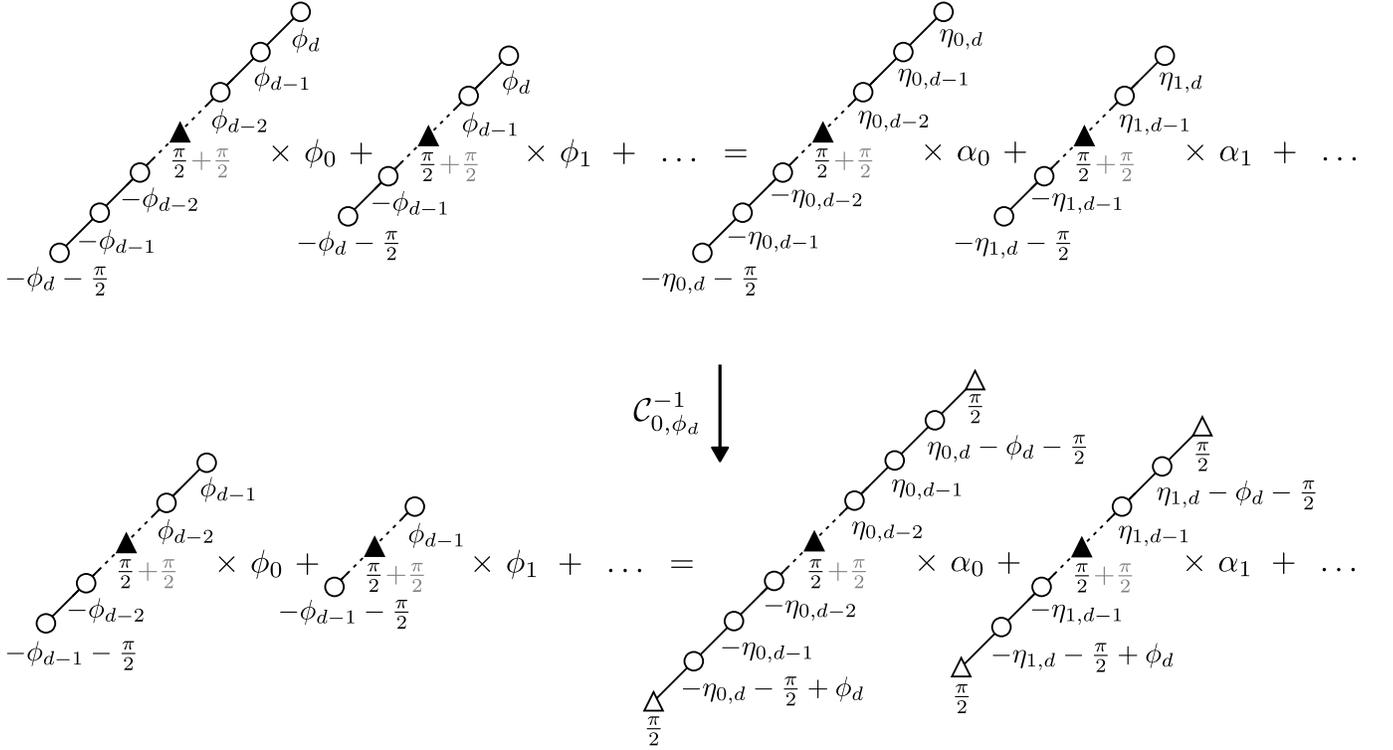

FIG. S7. Diagrammatic representation of one anti-conjugation by step in the proof of Theorem 4 for a length-$d$ QSP. The left-hand side depicts the first-order error terms and the right-hand side depicts a proposed set of recovery diagrams. Only the highest two degree terms in each sequence are shown as lower-degree diagrams cannot interfere assuming sufficiently large $d$ and $f(a) = O(1)$. The anti-conjugation by $\mathcal{C}_{0,\phi_d}^{-1}$ decreases the degree all terms of the error operator by 2 (save for degree-0 term which does not affect the analysis); In order for the right-hand side to match, it must be that $\eta_{i,d} = \phi_d$ for all recovery diagrams $i$.

for some $n_{d+1} \in \mathbb{Z}$ which implies $\psi_{d+1} - \phi_{d+1} = n_{d+1}\pi$. Furthermore, we find $\chi \in \{0, \pi\}$ using Remark S4.7. Thus proving the inductive step $\phi_{d+1} - \psi_{d+1} = 2\pi m_{d+1}$ and by extension, the Lemma. □

**Remark S6.2** (First-order error components are anti-Hermitian). Let $V$ be a first-order error component (i.e. of the form Definition S4.6 with all $b_i = 0$). First, note that is unitary since it is a QSP operator. Then from Corollary S6, we see that it is anti-Hermitian (i.e. $V^{-1} = V^\dagger = -V$).

**Remark S6.3** (Two error components cannot be combined, first-order). Let $U$ and $V$ be first-order error components of degree-$2r$ (i.e. of the form Definition S4.6 with all $b_i = 0$), parameterized by $\phi_{d-r+1}, \ldots, \phi_d$ and $\psi_{d-r+1}, \ldots, \psi_d$ respectively. Their weighted sum, $\alpha U + \beta V$ for $\alpha, \beta \in \mathbb{R}$ can be written as scaled single error component if and only if $\psi_i - \phi_i = n_i \pi$ for $n_i \in \mathbb{Z}$ for all $d - r + 1 \leq i \leq d$.

The $\impliedby$ direction follows directly from application of Remark S4.7. For the $\implies$ direction, consider that error components are QSP operators and therefore must be unitary. Therefore, we must have for some $c \in \mathbb{R}$,

$$(\alpha U + \beta V)(\alpha U + \beta V)^\dagger = (\alpha^2 + \beta^2)I + \alpha\beta(UV^\dagger + VU^\dagger), \tag{S120}$$

$$= (\alpha^2 + \beta^2)I - \alpha\beta(UV + VU), \tag{S121}$$

$$= cI, \tag{S122}$$

where we have used Remark S6.2 in the second equality.

Since $U$ and $V$ are of form Definition S4.6, their canonical expansions $\mathcal{P}$ and $\mathcal{P}'$ have $\mathcal{P}_j^{(0,1)} = \mathcal{P}_j'^{(0,1)} = 0$ for all $j$ by Remark S1.9. Thus, in order for the final equality in Eq. (S120) to hold, we must have $UV = VU = \pm I$ or equivalently $U^\dagger = -U = \pm V$. The result holds by application of Lemma S16.

We are now ready to prove Theorem 4 which for convenience is restated below.

**Theorem 4** (Lower bound on recovery length). *There exists a length-$d$ QSP sequence $U_\epsilon$ such that for any $XY$ recovery QSP $R_\epsilon$ of order $k \geq 1$ satisfying*

$$U_0^\dagger U_\epsilon R_\epsilon = I + \epsilon f(a) e^{i\frac{\pi}{2}Z} + O(\epsilon^2),$$

*for function $f(a) = O(a^0)$, $R_\epsilon$ has length $\Omega(d^2)$.*

*Proof.* Let $U_\epsilon = \mathrm{QSP}(\theta; (\phi_0, \ldots, \phi_d))$ be a noisy of length $d > 1$ QSP with error operator $E_\epsilon$ and $R_\epsilon$ any recovery sequence satisfying Remark S4.1 for $k \geq 1$.

From Remark S2.8, we see that each error component scaled by an independent real-value requires a separate irreducible recovery sequence of length $\Theta(d)$. To prove the Theorem, we show that generically $\Omega(d)$ independently scaled error components are required. We argue that to approximate the first-order error operator of Eq. (S37), we need as sequence of degree $2d, 2(d-1), 2(d-2), \ldots$ error components. In fact, given the restrictive condition of $f(a) = O(a^0)$, the only approximation is one that identical to the error operator up to the $\pi$ degrees of freedom allowed by Lemma S16.

Assume that we've found an approximation to first-order for an error operator of degree-$(2d)$. We proceed inductively, for the first $\Theta(d)$ diagrams by anti-conjugating thereby reducing the error operator to one of degree-$(2(d-1))$, neglecting the lowest-degree terms. Consider the first-order error terms of both $E_\epsilon$ and $R_\epsilon$ written in the form Definition S2.3. Anti-conjugating the error $\mathcal{C}_{0,\phi_d}^{-1} E_\epsilon$, results in all contributing diagrams decreasing in order by two (save for the degree-0 diagram) as the outermost phases of each diagram can be elided (Lemma S4). Therefore anti-conjugating the recovery operator $\mathcal{C}_{0,\phi_d}^{-1} R_\epsilon$ must likewise result in a two degree reduction. One step of the procedure is depicted in Fig. S7. Since by assumption, the difference in the $Z$-component is $f(a) = O(a^0)$, it cannot interfere with the top two degree diagrams for sufficiently large $d$; and the two highest-degree diagrams in $R$ must have outermost phase $\phi_d$ and be scaled by $\phi_0$ and $\phi_1$ respectively as in $E_\epsilon$. This can be seen by using Remark S6.1 and Remark S6.3. We can iterate this procedure $\Theta(d)$ times, before $f(a) = O(a^0)$ becomes relevant, each time requiring outermost phase of $\phi_{d-r+1}$ with scaling by $\phi_{r-1}$.

Thus the top $\Theta(d)$ degree diagrams in the recovery operator must be identical to that in the error operator. If all $\phi_i$ distinct, $\Theta(d)$ independently scaled error components are required, each of length $\Theta(d)$. Thus showing the lower bound of $\Omega(d^2)$ for general length-$d$ QSPs. $\square$

Our lower bound is independent of $k$ and is therefore loose for $k > 1$ and presents a direction for future work. An important limitation of our current construction is the recursive construction of higher-order recovery unites (Remark S4.12) which requires a doubling in length for each order in $k$. It remains an open question whether a order-$k$ unbiased sequence can be constructed using sub-exponential resources.

QSPs with phase degeneracies are able to circumvent this lower bound as in Remark S4.10. This motivates the exploration of families of polynomials that can be generated (or approximated) by QSPs with $o(d)$ unique phases.



\end{document}